\documentclass[a4 paper, 12pt]{article}
\usepackage[usenames,dvipsnames]{pstricks}
\usepackage[framemethod=tikz]{mdframed}
\usepackage[customcolors]{hf-tikz}
\usepackage[section]{placeins}
\usepackage[a4paper]{geometry}
\usepackage[toc,page]{appendix}
\usepackage[english]{babel}
\usepackage[T1]{fontenc}
\usepackage[utf8x]{inputenc}
\usepackage[flushleft]{threeparttable}
\usepackage{todonotes}
\PassOptionsToPackage{hyphens}{url}
\usepackage[bookmarksnumbered=true, bookmarksdepth=1,citecolor=blue, colorlinks=true, filecolor=blue,hyperfootnotes=false,linkcolor=blue, pdffitwindow=true,urlcolor=blue]{hyperref}
\usepackage{epsfig,etoolbox,graphicx,amsfonts,amssymb,amsmath,amsthm,xcolor,caption,framed,enumerate,footnote,marvosym,tabularx,booktabs,
ltxtable,setspace,rotating,eurosym, colortbl, bbm,pst-grad,pst-plot,natbib,longtable,lscape,subcaption,cleveref,epstopdf,upgreek,multirow,comment,lmodern,pbox,mathtools,mathrsfs}
\DeclareCaptionFormat{myformat}{#1#2\\#3}
\renewcommand{\thetable}{\arabic{table}}
\renewcommand{\thefigure}{\arabic{figure}}

\usepackage{tikz}
\usepackage{tikz-cd}
\usetikzlibrary{shapes,arrows,shadows,positioning,calc}

\usepackage{dcolumn}
\newcolumntype{d}[1]{D{.}{.}{#1}}

\usepackage{float}
\usepackage{courier}
\usepackage{soul}

\newcommand{\sym}[1]{\rlap{#1}}

\usepackage{siunitx} 
\newcolumntype{^}{>{\currentrowstyle}}

\begin{document}

\title{\vspace{-1cm}{\bf Prenatal Sugar Consumption and Late-Life \\ Human Capital and Health: 
Analyses Based \\ on Postwar Rationing and Polygenic Scores}\vspace{0.5cm}}
\author{Gerard J. van den Berg\thanks{ University of Groningen, University Medical Center Groningen, IFAU Uppsala, IZA, ZEW, J-PAL and CEPR.} \and Stephanie von Hinke\thanks{ Corresponding author: School of Economics, University of Bristol; Institute for Fiscal Studies. E-mail: \href{mailto:S.vonHinke@bristol.ac.uk}{S.vonHinke@bristol.ac.uk}} \and R. Adele H. Wang\thanks{School of Economics, University of Bristol.
\newline \newline
We thank Andreas Peichl, Pietro Biroli, Fleur Meddens, Hans van Kippersluis, Niels Rietveld and participants at a seminar at the Erasmus School of Economics and a CESifo/NORFACE conference for useful comments and discussions. Samuel Baker, Sean Harrison, and Emil N\o{}rmark S\o{}rensen provided valuable help with the data. We gratefully acknowledge financial support from NORFACE DIAL (Grant Reference 462-16-100) and the European Research Council (DONNI 851725).}\vspace{0.5cm}}

\date{\textit{This version of the paper first circulated as a working paper in} \\
\medskip 
September 2022}

\maketitle

\begin{abstract}
\noindent Maternal sugar consumption in utero may have a variety of effects on offspring. We exploit the abolishment of the rationing of sweet confectionery in the UK on April 24, 1949, and its subsequent reintroduction some months later, in an era of otherwise uninterrupted rationing of confectionery (1942-1953), sugar (1940-1953) and many other foods, and we consider effects on late-life cardiovascular disease, BMI, height, type-2 diabetes and the intake of sugar, fat and carbohydrates, as well as cognitive outcomes and birth weight. We use individual-level data from the UK Biobank for cohorts born between April 1947--May 1952. We also explore whether one's genetic predisposition to the outcome can moderate the effects of prenatal sugar exposure. We find that prenatal exposure to derationing increases education and reduces BMI and sugar consumption at higher ages, in line with the ``developmental origins'' explanatory framework, and that the sugar effects are stronger for those who are genetically predisposed to sugar consumption. 

\vspace{3mm}

\vspace{1mm}\newline
\end{abstract}

\newpage


\section{Introduction}\label{sec:intro}

Sugar has been part of the human diet for centuries, providing energy and making food more palatable. More recently, sugar consumption has been related to a range of diet-related health problems. WHO (2015) recommends that the individual intake of sugars is restricted to less than 10\% of the individual's total energy intake, with the following motivations: (i) to prevent obesity, (ii) to prevent that sugar-laden foods substitute more healthy dietary items, and (iii) to prevent caries. Both (i) and (ii) are thought to mitigate the risks of cardiovascular disease, type-2 diabetes and other diet-related diseases. 

These health concerns are backed up by countless studies of adults and children, documenting effects that materialize in the short to medium run. Much less is known about effects of exposure to sugars in utero and in particular, about long-run hazards of such exposure on health at advanced ages or on human capital outcomes. Following the literature on the developmental origins of late-life health, one may hypothesize the existence of in utero exposure effects. At first sight, this literature seems to suggest that a high intake helps to prevent undernutrition and provides an in utero environment that is aligned with later-life nutritional conditions, improving late-life health.\footnote{See e.g., Almond and Currie (2011), Lindeboom et al. (2010), Roseboom et al. (2006), Scholte et al. (2015), and Bengtsson and Lindstrom (2000, 2003) for overviews and studies of the causal evidence of effects of in utero nutrition on late-life health and economic outcomes.}  However, this literature does not analyze effects of sugar in isolation from other nutrients and hence it ignores the specific nature of sugar as a nutrient with problematic aspects. Sugar exposure may adversely affect the build-up of the body, leading to worse late-life health. This second pathway is supported by the fact that the set of late-life health outcomes that are known to be responsive to adverse in utero conditions in general (such as cardiovascular diseases, type-2 diabetes and cognitive decline) closely resembles the set of health outcomes responsive to adult abundant consumption of sugars. In either case, knowing whether causal long-run effects exist is vastly important for public policy.

A number of descriptive studies have examined the association between maternal gestational sugar consumption and offspring early-life health outcomes such as birth weight, childhood cognition and obesity. A higher intake of sugars (e.g. as sweets, snacks and soft drinks) is often found to be associated with less healthy outcomes although the evidence is not unanimous.\footnote{See e.g. Walsh et al. (2011) for obesity, Phelan et al. (2011) for birth weight and Cohen et al. (2018) for childhood cognition. B\'edard et al. (2017) find an association with atopy and asthma in later childhood. See Goran et al. (2018) and Casas et al. (2020) for reviews of existing studies.} These studies, however, do not consider late-life outcomes, and, crucially, they do not control for the endogeneity of sugar consumption.\footnote{Controlled laboratory experiments with rodents have demonstrated adverse effects of in utero sugar exposure on offspring's adult obesity; see e.g. Bocarsly et al. (2012).}

Clearly, in general, parents choose their sugar consumption levels, and as a result, these levels will depend on a wide range of characteristics of the parents and their circumstances. Many of those characteristics also affect other choices and behaviors, thus influencing the outcomes of their offspring. Any association between sugar intake and offspring health may therefore reflect shared unobserved determinants. To determine whether the associations also reflect causal effects calls for causal inference. 

The present paper uses a causal inference approach to study effects of in utero exposure to sugar on late-life health and human capital outcomes. We exploit features of the food rationing system in the UK in the years immediately following World War II. Specifically, we exploit the abolition of the rationing of confectionery (including chocolate and sweets) in the UK on April 24, 1949 and its subsequent reintroduction on August 13, 1949. This ``derationing period'' occurred in an era of otherwise uninterrupted rationing of confectionery (1942-1953) and of raw sugar (1940-1953) as well as many other foods. In a nutshell, we compare individuals who were in utero during the short derationing period to those who were in utero just before or after this period.

For our purposes, it is an advantage that the derationing in April 1949 and its reversal in August 1949 took place in a period without other rationing abolitions or introductions, neither of sugar nor of other consumption goods (see Zweiniger-Bargielowska, 2000; see also \hyperref[sec:AppendixA]{\ref*{sec:AppendixA}} of our paper). The vast majority of rationing measures was introduced before mid-1942, and most ended in the 1950s. The derationing interval in 1949 did not witness other major societal events either. The April 1949 derationing itself was imposed through an administrative act taken by civil servants in response to pressure by the confectionery industry. Importantly, therefore, the decision was not taken by the cabinet which could have used it to curry favor with the electorate and which often discussed minor changes in ration levels. The subsequent (unplanned) reintroduction in August 1949 was due to the inability of the confectionery industry to meet the demand for its products, which in turn was due to the unwillingness of the government to allocate or transfer larger amounts of sugar to that industry, coupled with a lack of willingness to allow for market price mechanisms (see Zweiniger-Bargielowska, 2000, for background information on the derationing and its reversal).\footnote{Other changes in rationing took place in time periods that witnessed additional upheavals, making it more challenging to identify their separate contributions to health and human capital outcomes. Notably, the onset of rationing coincided with preparations for the war, while the ultimate derationing of confectionery and sugar in 1953 took place in a period in which the rationing of many other products was abolished as well (see \hyperref[sec:AppendixA]{\ref*{sec:AppendixA}}) and in which the economy moved from a socialist design to a free-market design following a right-wing election victory. A notable event close to our derationing interval is the formal establishment of the NHS universal health service in England in July 1948. However, the actual health care system and supply changed rather gradually in this period (see Rivett, 2014, for a comprehensive overview, and Thomson, 2018). Many individuals were insured before July 1948 and aspects of universal care were already available regionally, while after July 1948 certain provisions were in short supply and led to queueing, and other provisions remained costly. The year 1948 also witnessed a gradual tightening of educational requirements for nurses. These overall changes in health care can therefore be accounted for by a flexible trend specification in calendar time in our empirical analysis.}

The reversal of the derationing facilitates the identification of gestational ages with high exposure. By analogy to the Dutch Hunger Winter studies, the identification exploits that the exposure period is substantially shorter than the length of gestation (see Lumey et al., 2011, and Scholte et al., 2015). Study designs based on a single discontinuity do not have this advantage.

Notice also that the maternal dietary changes induced by the brief derationing are relatively mild compared to the major shocks (such as famines) that are commonly exploited in the causal inference literature on conditions in utero. This makes our analysis less sensitive to selective fertility issues that have plagued the latter literature. In addition, sugar and confectionery consumption is amenable to specific intervenable policy targets, whereas nutritional policy motivated by famine studies requires extrapolations from extreme events to modest interventions. 

An obvious challenge to the usage of rationing as a proxy for sugar and confectionery consumption is that individuals may adjust their consumption of other foods in response to rationing. First, confectionery may be consumed in combination with certain other nutrients, such as coffee. However, the widespread rationing of other products means that the derationing of confectionery is unlikely to have led to a change in the consumption of those nutrients, simply because the latter faced an ongoing binding constraint. Secondly, during rationing, sugar may be substituted by replacement products. In the rationing era this included saccharin as well as barley extract and liquorice (see Royle, 2020). This is relevant for our study design insofar that such substitutes may affect health at higher ages. In their overviews, Goran et al. (2018) and Casas et al. (2020) report evidence of a positive association between the gestational intake of artificial sweeteners and infant and child obesity (as well as an increased preference for sweet tastes) and they conjecture that this is explained by effects of sweeteners on the gut microbiome of the child. It is not clear if this extends to late-life health, but Palatnik et al. (2020) report long-run effects along these lines in animal studies. 

Another challenge concerns effects of exposure to sugar and confectionery after birth. In our setting, children born before the derationing period in mid-1949 are exposed during some months in 1949. Moreover, every child born before the ultimate permanent derationing in 1953 is exposed from 1953 onwards. The literature on long-run effects of early-life conditions cited above generally finds that effects of in utero exposure exceed those of exposure after birth. In our paper we estimate a variety of model specifications and we use various subsamples of individuals born before, during and after the derationing period in mid-1949. We also point out that exposures after 1953 are fully in line with our aim to study effects of prenatal exposure for those who, in later childhood and in adulthood up to high ages, live in a society with abundant nutrition. We return to this below.

The data set we use is the UK Biobank which follows the health and well-being of approximately 500,000 individuals in the UK aged 40-69 between 2006-2010 (see Sudlow et al., 2015). All participants are born between 1934 and 1971, and the vast majority belongs to the 1940s-1960s birth cohorts. UK Biobank participants have provided survey information on their health and well-being and have given blood, urine and saliva samples, and all individuals have been genotyped. The individuals exposed in utero to derationing were on average approximately 60 years old upon entering the Biobank.
As motivated by the literature, our set of late-life outcome variables includes educational attainment, cardiovascular disease, BMI, height, type-2 diabetes, and the intake of sugar, fat and carbohydrates. We also examine effects on birth weight.\footnote{Birth weight has been used as a biomarker of conditions in utero although recently doubt has been cast on the extent to which it reflects conditions that are relevant for late-life health (Schulz, 2010; Van den Berg and Modin, 2013; Maruyama and Heinesen, 2020). An obvious advantage of birth weight as an outcome measure is that it does not depend on exposure after birth.} 

Our paper provides two sets of major and novel findings. First, these concern the direct effects of prenatal exposure to sugar derationing. We find that the latter improves education and reduces BMI and sugar intake later in life, with some evidence that it also led to an increase in birth weight. 
Second, we explore whether one's genetic predisposition can moderate the effects of prenatal exposure on later life outcomes. Historically, the study of gene-by-environment ($G \times E$) interactions is hampered by the possible endogeneity of the environment, as an individual's genetic predisposition may drive them to self-select, seek out or create certain environments (so-called gene-environment correlation, or $rGE$; Plomin and Bergeman, 1991). Natural experiments such as the exogenous exposure to temporary derationing can provide a solution to this issue. This makes our paper one of the first of only a handful of recent studies to analyze $G \times E$ interactions within a natural-experiment setting, enabling us to make causal inferences on the nature-nurture interplay.\footnote{Previous examples include Fletcher (2012, 2019), Schmitz and Conley (2016), Barcellos, Carvalho and Turley (2018), Biroli and Z\"und (2020), Pereira et al. (2020) and Muslimova et al. (2020). A concurrent example is von Hinke and S{\o}rensen (2022).}  

We use individuals' so-called polygenic scores (also known as polygenic indices) for the respective health and education outcomes as measures of the genetic predisposition for these traits. Classical twin studies indicate that the heritability\footnote{Heritability is the proportion of the variance in a trait that can be attributed to genetic differences between people in a population.} of the health outcomes ranges between 30\% and 72\% (see e.g., Avery and Duncan, 2019; Maes, Neale and Eaves, 1997; Marenberg et al., 1994) while the heritability of educational attainment ranges from 25\% to 40\% (see e.g., Branigan, McCallum and Freese, 2013; Lucchini, Della Bella and Pisati, 2013). This suggests that variation in these traits is under at least a moderate degree of genetic influence. The recent development of large-scale genome-wide association studies (GWAS) has allowed for the identification of specific genetic variation, namely single nucleotide polymorphisms (SNPs), associated with traits of interest.\footnote{See \hyperref[sec:AppendixB]{\ref*{sec:AppendixB}} for an introduction to the genetic concepts used here.} The results of these studies allow us to construct polygenic scores for each individual, quantifying one's genetic ``predisposition'' to each of the outcomes. For our purposes, these can be straightforwardly interacted with exogenous changes in the prenatal sugar environment.

We find some evidence of genetic heterogeneity of effects of exposure. More specifically, we find that high polygenic scores for cardiovascular disease and for sugar and carbohydrate intake affect the impact of a high-sugar prenatal environment. Interestingly, those who are exposed to a high-sugar prenatal environment reduce their late-life sugar intake more strongly if they happen to be genetically predisposed to the consumption of sugar in adulthood. Relating this to the ``developmental origins of late-life health'' framework, this suggests that  the protective effect of fetal programming is stronger for those who are genetically predisposed to sugar consumption. 

The rest of the paper is structured as follows. Section 2 describes wartime and post-wartime rationing in the UK in general, and for sugar and confectionery in particular. Section 3 discusses the data. Section 4 describes the empirical specification and Section 5 presents the results. Section 6 explores the robustness of our findings by providing a range of sensitivity analyses. This includes alternatives for the length of the exposure interval in 1949, the birth cohorts used, the specification for the trend effect of calendar time at birth on the outcomes, analyses by gender, and specificities of the exposure by gestational trimester. Section 7 concludes.


\section{Rationing in the UK during and after World War II}
\label{sec:Background}

\subsection{The rationing system}

On January 8, 1940, four months after the start of World War II, the UK wartime government introduced food rationing.\footnote{This section draws heavily on Zweiniger-Bargielowska, 2000, and references therein.} At the time, for many food categories, the UK imported almost everything it consumed. With an envisaged significant drop in food imports, the government expected severe food shortages, rising prices, and a growing inability of low SES households to afford food. In addition, food availability was expected to be volatile, leading to hoarding. By analogy to the experiences in Germany towards the end of World War I, food shortages were expected to affect the morale and lead to food riots and civil unrest. The main aim of food rationing was therefore to ensure that everyone had access to necessary foods. Incidentally, the UK rationing system did not only cover food but also clothing, fuel, soap, paper, furniture and other products. Furthermore, the rationing system was not only targeted at individual consumers but, depending on the product involved, also restricted inputs of industries, sectors and public services. 

To purchase rationed foods, individuals were issued ration books, with different coupons for different foods, e.g., separate coupons for butter, cheese, and sugar. Ration coupons were the government guarantee that you would be able to get your share of the food, though it could only be used at the shop where the individual had registered, with shop keepers cutting out the relevant coupons when individuals purchased the product.\footnote{In December 1941, so-called ``point rationing'' was introduced, allowing some freedom to choose both the product (mainly canned food) and shops. 
There was a very small black market in luxury goods; a black market in necessities hardly existed (Ministry of Health, 1946).} \hyperref[sec:AppendixA]{\ref*{sec:AppendixA}} provides a table with rationed food types and the time intervals in which each of these was rationed. 

Throughout the rationing era, pregnant women, nursing mothers and children under five were entitled to special rations for milk, eggs, meat, orange juice, cod liver oil, and vitamin A and D (Davies, 2014). By the end of the war, most schools served school dinners, which were either free or at cost-price. Furthermore, it was compulsory for any factory employing over 200 workers to open a canteen to provide meals to its workers. These did not require usage of coupons.

As a secondary aim, the rationing system was used to improve the quality of individuals' nutritional intake and reduce nutritional inequalities between SES groups by steering the accessible food bundle towards a nutritionally healthy diet available to all. The evidence shows that this was successful (see Ministry of Health, 1946; Burnett, 1987; Department of Health, 2002; Goldring et al., 2011).\footnote{Intakes of protein, calcium, iron, vitamin B, riboflavin, vitamin C, and nicotinic acid all increased considerably during wartime rationing. This meant that the average diet of all classes was better balanced than ever before (Ministry of Health, 1946).} Children were taller and heavier than before the war, infant mortality rates reduced and maternal mortality rates dropped (Ministry of Health, 1946; Burnett, 1987). 

The years immediately following the end of World War II witnessed a global food crisis, while the UK continued to be heavily dependent on food imports. The war had taken a heavy toll on the UK national debt and the UK insisted on maintaining a large military force abroad after the war. As a result, it faced budgetary and balance of payments difficulties. The postwar socialist government saw food rationing as a policy to limit consumption of imported goods without putting poor households in jeopardy. Hence, it decided to continue the food rationing system and to even further extend the types of food products subject to rationing. This policy continued into the early 1950s when the socialist government lost the national election. Note that the postwar rationing exploited in our study design took place in a setting that was less dramatic than rationing in war-related famines (see e.g. van den Berg et al., 2016).

\subsection{Rationing of sugar and confectionery} 

Before the war, about 70\% of the sugar used in the UK was imported. Sugar was one of the first batches of products to be rationed in early 1940 and among the last to be derationed in late 1953. Importantly for our purposes, sugar rationing was regarded to be more binding and restrictive than the rationing of most other food products, in the sense that its consumption was reduced more dramatically (see Zweiniger-Bargielowska, 2000, and references therein). 

By June 1942, UK food consumption had become heavily dependent on imports from the US and other allies, and food trade was coordinated at the international level. In the first half of 1942, the US lost the Philippines (its prime supplier of sugar) to Japan (see e.g. Keeton, 2011), and import from other areas was hampered by sea warfare. This led to the introduction of sugar rationing in the US, which was followed by the rationing of confectionery products in the UK on July 26, 1942. The latter adopted the ``point rationing'' system with \textit{individual} (rather than \textit{household}) allowances of $\approx$3 oz (85g) per person per week.\footnote{The exact amount varied slightly throughout the rationing years depending on the supply of raw ingredients and special events, see \autoref{fig:rations}. For example, those under 18 were given an additional 2 oz ``Christmas bonus'' in December 1944 and December 1946 (MAF 156/262, 1947). Sugar and confectionery are not identical but confectionery mostly consists of sugar. Additional ingredients may include small amounts of fat, flour and proteins. Chocolate and sugar confectionery were items that could be purchased with the relevant ration coupons. These two items were consumed in approximately equal proportions in the pre-war years, but the rationing era saw a preference for chocolate with occasional minor surpluses of sugar confectionery, which was subsequently used by the confectionery industry as an argument for the abolishment of rationing in 1949 (Financial Times, 15/02/1949).} 

As we alluded to in Section 1, pressure from the industry and an improvement in the supplies of sugar and other ingredients made the relevant ministry decide to completely deration confectionery starting on April 24, 1949. The announcement was made on February 21, 1949 (Financial Times, 22/02/1949). The industry estimated that the supply of raw materials at the time would allow weekly consumption per capita to rise to 4.75 oz (135g). The trade bodies estimated that demand and supply would find an equilibrium at this level without price adjustments which would have been politically infeasible (Financial Times, 22/02/1949). To allow traders to build up stock and meet the impact of derationing, civil servants allocated an immediate 50\% additional credits to sweets wholesalers and retailers between February and April 1949, plus a further one-off grant of 50\% of the coupons collected in March 1949 (MAF 156/270, 1952). 

As it happened, the actual demand of 8-10 oz (227-283 g) far outstripped supply, with sales in late April 1949 exceeding the levels in April 1948 by 84\% (Financial Times, 06/11/1949), leading to long queues at confectionery shops, and a rapid running down of any stocks (see Zweiniger-Bargielowska, 2000; Royle, 2020). By late May 1949, many confectionery shops had run out of stock and closed for business (Fitzgerald, 1995). In a nutshell, the confectionery industry was caught in-between the lack of inputs and the fact that consumer product prices were fixed.\footnote{Other industries also faced restrictions due to sugar rationing; for example, beer breweries had to modify the composition of their products.} In addition, export of confectionery production had priority over domestic consumption (Royle, 2020). With the Chancellor of the Exchequer announcing a cut in sugar imports for July, the government decided on July 14 to re-introduce confectionery rationing on August 13, 1949; less than four months after the initial derationing. Final derationing of confectionery and sugar only took place in February and September of 1953, respectively, along with the abolishment of rationing of most other products. Subsequent years witnessed steeply increasing consumption.

The above implies that the derationing on April 24, 1949 was exogenous, precisely delineated in time, and difficult to anticipate at the individual level. The reinstallment of rationing was also exogenous at the individual level. However, the point in time at which individuals started to experience confectionery purchase constraints (due to shops running out of stock) most likely took place before the official re-instalment of rationing on August 13, 1949. Therefore we perform a sensitivity analysis, stipulating derationing to end on May 31, 1949.


\section{Data}
\label{sec:Data}

The UK Biobank (UKB) is a prospective, population-based cohort study that contains data on the health and wellbeing of over 500,000 individuals across the United Kingdom. Recruitment and collection of baseline data took place between 2006 and 2010, when participants were aged between 40-69. Individuals provided information on demographics, health status, lifestyle measures, cognitive testing, personality self-report, and physical and mental health measures via questionnaires and interviews. Furthermore, anthropometric measures were taken by nurses, along with blood pressure readings and samples of blood, urine and saliva.  All individuals have been genotyped.

We restrict our sample in three ways. First, we only consider those born between April 24, 1947 and May 13, 1952 (i.e., two years prior to the start of derationing and two years after the last individuals exposed to derationing in utero were born). This restricts the sample to 106,608 individuals. Second, as individuals' genetic architecture is known to differ across ancestry groups, we follow the literature and exclude non-whites. Third, we drop individuals for whom no geographical birth coordinates are available or who were born in Scotland or Northern Ireland as region of birth is not available for those parts of the UK. This leads to dropping an additional 15,221 individuals. Our final sample includes a maximum of 84,539 individuals, depending on the outcome of interest.

We select a set of nine outcomes as motivated by the literature on long-term effects of poor prenatal environments (see Section 1) and their availability in the UK Biobank. We create a measure of years of education based on the International Standard Classification of Education (ISCED) scale, along the lines of the literature (see e.g. Rietveld et al., 2013; Okbay et al., 2016, 2022; Lee et al., 2018). Birth weight is self-reported in the UK Biobank, but it is considered to be reliable in that it correlates with variables that are known to correlate with true birth weight, such as gender, ethnicity and maternal smoking (Horikoshi et al., 2016; Tyrrell et al., 2013). BMI is based on in-clinic measurements of height and weight. A binary indicator for cardiovascular disease is obtained from primary and secondary diagnoses codes (ICD-10) in participants' hospital inpatient records that are linked to the UK Biobank, or via the cause of death that is available in the National Death Registries. Type-2 diabetes is self-reported by the participant. Finally, macro-nutrient intakes represent nutrient \textit{densities}; i.e., the proportion of the total energy intake that is due to sugar, carbohydrates and fats (see e.g., Willett et al., 1997). 

In addition to the extensive socio-demographic and health information observed in the UK Biobank, all participants have been genotyped. We use this genetic information to create ``polygenic scores'': index variables measuring individuals' genetic predisposition towards a specific trait. In short, a polygenic score is a weighted sum of the individual SNP effects (see \hyperref[sec:AppendixB]{\ref*{sec:AppendixB}} for a more detailed discussion), defined as: 

\begin{equation}\label{eq:PGS}
G_i = \sum_{j=1}^J \widehat{\beta}_j X_{ij}
\end{equation}

\noindent
where $G_i$ is the polygenic score of a specific trait for individual $i$, $X_{ij} \in {0,1,2}$ is a count of the number of ``risk'' alleles for individual $i$ at SNP $j$, and $\widehat{\beta}_j$ is a weight associated with SNP $j$. These weights are obtained from independent genome-wide association studies (GWAS). Using a GWAS sample that is more similar to the analysis sample increases predictive power, since the polygenic score captures the genetic predisposition within the environmental context of individuals from the discovery sample (Domingue et al., 2020). To construct the polygenic scores used here, we therefore run our own tailor-made GWAS on UKB participants, excluding our analysis sample (i.e., those born between April 24, 1947 and May 13, 1952) to avoid over-fitting, and we drop any siblings of UKB participants to avoid clustering. We then use the summary statistics from this GWAS to create the relevant polygenic scores for our analysis sample, using LDpred (Vilhj{\'a}lmsson et al., 2015; see \hyperref[sec:AppendixB]{\ref*{sec:AppendixB}} for more information on how the polygenic scores are constructed). All polygenic scores are standardized to have mean zero and standard deviation one.

\autoref{tab:descr} presents the descriptive statistics of the outcomes of interest, including the maximum sample sizes for the analysis sample of each outcome. This shows that, on average, UKB participants born between April 1947 and May 1952 have almost 14 years of education, a birth weight of 3332g and a BMI of 27.6. Approximately 33\% of the sample has been diagnosed with cardiovascular disease and 5\% have been diagnosed with type-2 diabetes. Approximately 23\% of individuals' total energy intake is derived from sugar, 48\% from carbohydrates and 33\% from fats. Just under half of the sample is male.

\begin{center}
	\textbf{[\autoref{tab:descr} about here]}
\end{center}

\begin{figure}[H]
\caption{Trends in the outcomes of interest}
\label{fig:Trends}
\centering
\begin{subfigure}{.4\linewidth}
  \centering
  \includegraphics[width=1\linewidth]{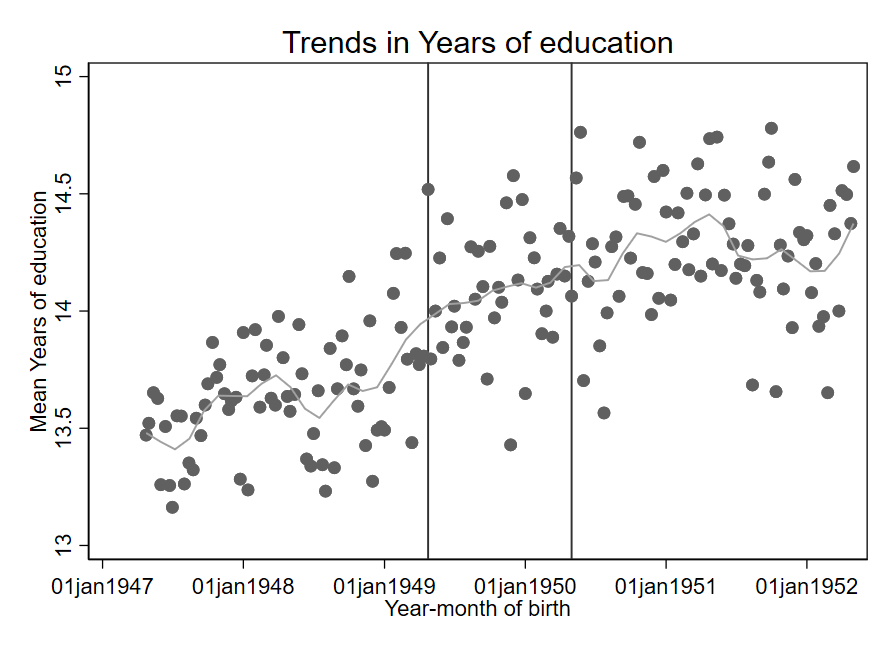}
  \caption*{}
\end{subfigure}
\begin{subfigure}{.4\linewidth}
  \centering
  \includegraphics[width=1\linewidth]{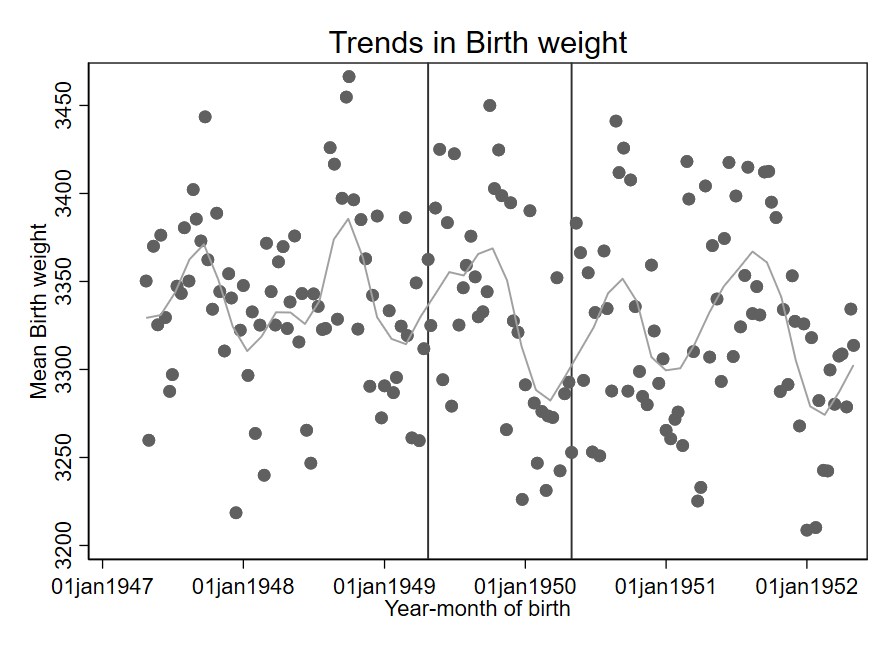}
  \caption*{}
\end{subfigure}\\[1ex]
\begin{subfigure}{.4\linewidth}
  \centering
  \includegraphics[width=1\linewidth]{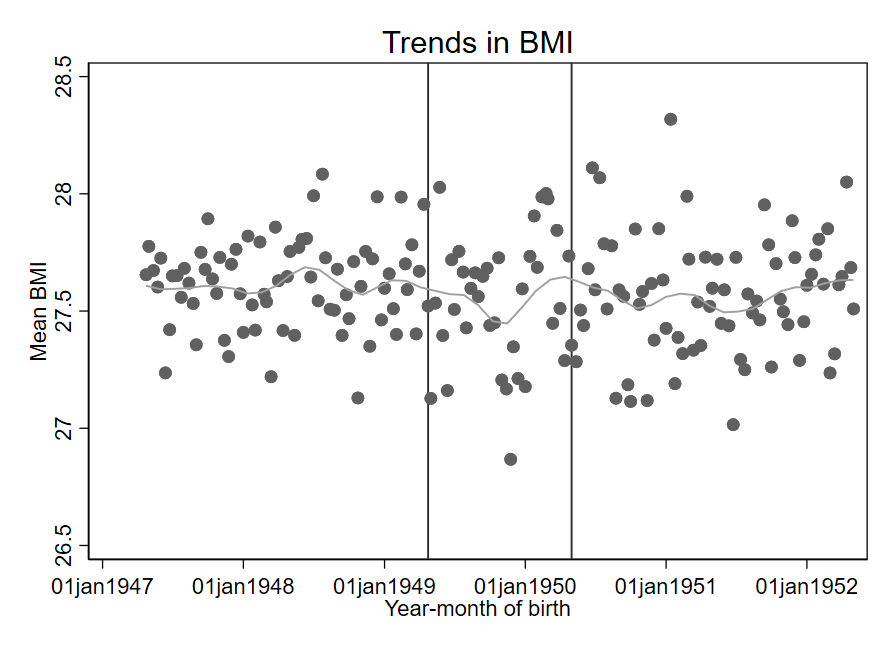}
  \caption*{}
\end{subfigure}
\begin{subfigure}{.4\linewidth}
  \centering
  \includegraphics[width=1\linewidth]{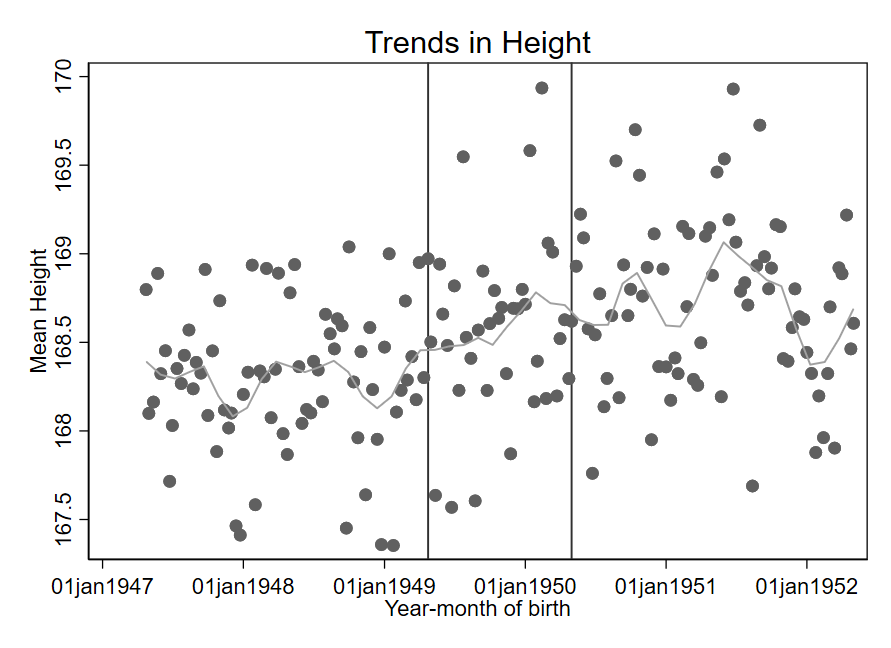}
  \caption*{}
\end{subfigure}\\[1ex]
\begin{subfigure}{.4\linewidth}
  \centering
  \includegraphics[width=1\linewidth]{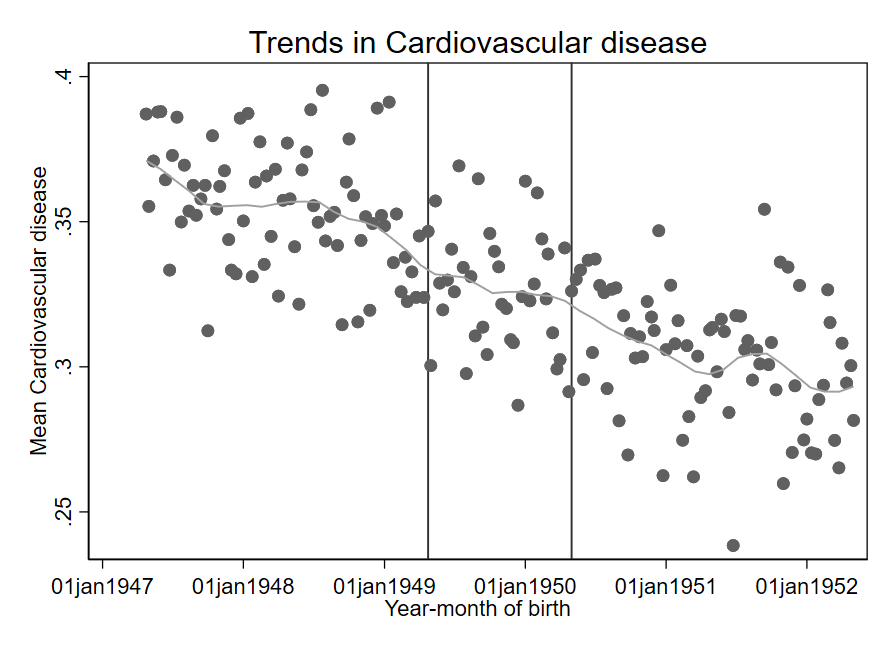}
  \caption*{}
\end{subfigure}
\begin{subfigure}{.4\linewidth}
  \centering
  \includegraphics[width=1\linewidth]{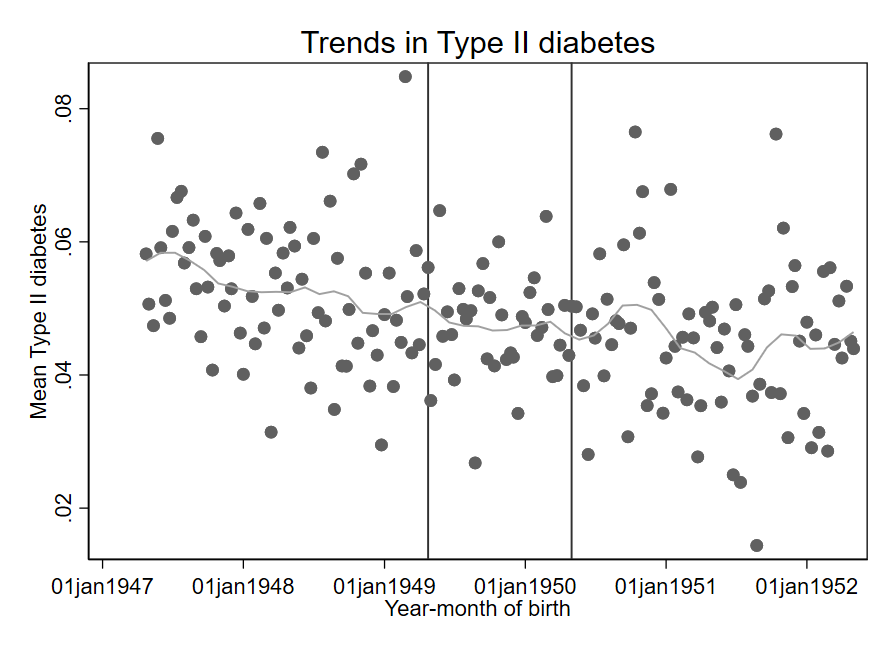}
  \caption*{}
\end{subfigure}\\[1ex]
\begin{subfigure}{.4\linewidth}
  \centering
  \includegraphics[width=1\linewidth]{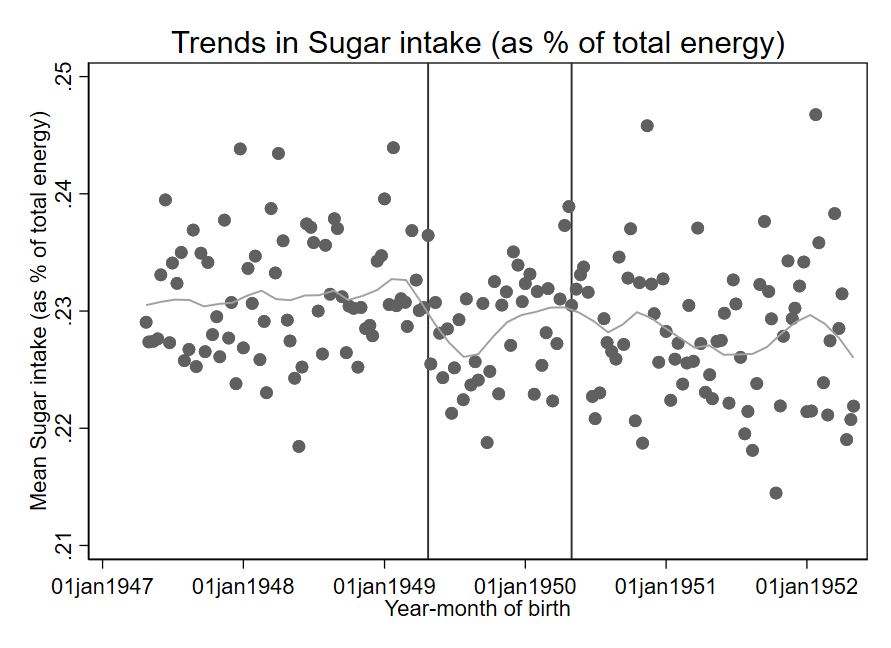}
  \caption*{}
\end{subfigure}
\begin{subfigure}{.4\linewidth}
  \centering
  \includegraphics[width=1\linewidth]{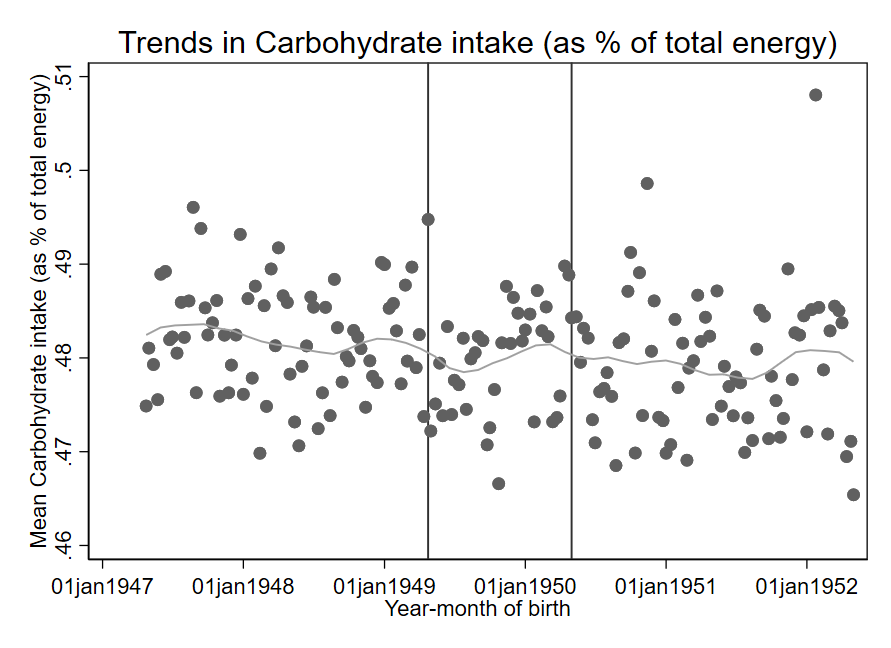}
  \caption*{}
\end{subfigure}\\[1ex]
\caption*{\scriptsize{Notes: The figures plot the trends of the relevant outcome variable by year-month of birth.}}
\end{figure}

\autoref{fig:Trends} shows the trends in each outcome variable for our period of observation by individuals' year-month of birth, where the two vertical lines indicate the first and last cohort exposed prenatally to the temporary confectionery derationing period.\footnote{Henceforth, ``year-month'' is used to denote the combination of calendar year and calendar month, e.g. April 1949. To create \autoref{fig:Trends}, we separate each month into ``early'' (born prior to the 12th of the month), ``mid'' (between the 13th  and 23rd of the month) and ``late'' (born on the 24th of the month or after). Hence, we plot three observations for each month. The exposed cohorts are born between April 24, 1949 and May 12, 1950 (i.e., 9 months (minus one day) after the reintroduction of rationing on August 13, 1949).} The figures suggest a slight reduction in BMI and sugar intakes for those prenatally exposed to the derationing. However, there is substantial variation in each of the outcome variables. We explore the data more rigorously below.

\begin{center}
	\textbf{[\autoref{fig:Trends} about here]}
\end{center}

To gauge evidence for gene-environment correlations ($rGE$), \autoref{fig:rGE} in \hyperref[sec:AppendixC]{\ref*{sec:AppendixC}} plots the densities of each of the (standardized) polygenic scores for those exposed to derationing in utero (solid line) and those unexposed (dashed line). From the very similar polygenic score distributions we conclude that there is little evidence of polygenic scores being correlated with the exposure indicator. The scatter plots overlaying the densities are the averages of each outcome for 200 equal sized bins of the corresponding polygenic score. The fitted curve through the dots is obtained from a kernel-weighted local polynomial regression of the outcome on the corresponding polygenic score, where we again distinguish between the exposed and unexposed groups. This shows a strong correlation between the polygenic score and each of the corresponding outcomes. The statistical strength of this relationship is investigated below. Finally, the figures suggest that we can approximate the relationship between the polygenic score and the respective outcome as linear.


\section{Empirical approach}
\label{sec:Methods}

The baseline specification for the empirical analysis is: 

\begin{equation}
\label{eq:E}
y_i = \alpha_0 + \alpha_1 E_i + \delta X_i + f(X_i,E_i) + \varepsilon_i 
\end{equation}

\noindent
where $y_i$ is the outcome of interest for individual $i$ and $E_i$ is a binary indicator equal to one if the individual was exposed to at least one day of confectionery derationing whilst in utero (i.e., was born between April 24, 1949 and May 12, 1950).\footnote{In the absence of data on gestational age, we here assume that all pregnancies last 9 months.} Hence, our coefficient of interest is $\alpha_1$, capturing the effect of prenatal exposure to confectionery derationing on the outcome of interest. $X_i$ includes gender and calendar time by way of a (in the baseline specification) linear function of the year-month of birth. It also includes 11 binary indicators for the month of birth to remove seasonality in outcomes. The function $f(X_i,E_i)$ includes interactions between the covariates $X_i$ and our ``exposure indicator'' $E_i$.\footnote{This only includes the interaction of $E_i$ with calendar time and with gender. We add the $f(X_i,E_i)$ interactions to allow for comparisons of the estimates with those from \autoref{eq:GxE} below, though our estimates are robust to not including any such interactions. We do not include interactions with the month fixed effects since its coefficient would partially capture the effect of interest. For example, the interaction between the month of birth February and $E_i$ captures those who were in utero at some point between April 24, 1949 and May 12, 1950 and were born in February. By construction, this only includes all individuals born in February 1950. Since this is part of our exposure effect, the inclusion of that interaction would inadvertently capture that.} Finally, $\varepsilon_i$ captures omitted determinants of the outcome as well as measurement error in the outcome. We report heteroskedasticity-robust standard errors clustered by year-month of birth.

Note that the above approach constitutes an ``Intention To Treat'' (ITT) design for the evaluation of actual intake, as we do not observe the individual consumption of confectionery.\footnote{Historical documents and newspaper coverage document that the demand for confectionery rose dramatically during the derationing period compared to the period before. Due to a lack of data on actual confectionery \textit{consumption}, however, we cannot estimate (or instrument) the prenatal \textit{consumption} effect on offspring outcomes. The National Food Survey, the main UK survey of household food consumption that started in the 1940s, does not record intake of chocolates and sugar confectionery for this period since it is considered a \textit{personal} rather than \textit{household} good (Ministry of Food, 1951).} 
Given that the derationing was announced only two months before its implementation, we assume that the timing of derationing relative to pregnancies is exogenous. We examine this empirically in more detail in a robustness analysis, where we also estimate model varieties with a more narrow exposure interval.

There is geographical variation in the nutritional environment during the period of rationing. For example, it is plausible that eggs, butter and meat were more available in rural areas where individuals had more access to home-grown foods. Similarly, the demand for confectionery was higher in cities and there is evidence that confectionery shops in city centres were among the first to run out of stocks (Fitzgerald, 1995). To account for differential local supply and demand of confectionery, we re-estimate equation (\ref{eq:E}) including region-of-birth fixed effects. During the period of observation, England and Wales were divided into over 60 counties, 230 administrative counties, and 1400 local government districts (henceforth: districts). Using information on individuals' location of birth, we assign to each individual a county, administrative county and district ID. Including such fixed effects enables us to compare individuals in close proximity but at different points in time, some of whom would have been exposed to prenatal derationing and others not. In these specifications, we cluster our standard errors by region of birth.

The estimates of \autoref{eq:E} capture the average effect of prenatal exposure to confectionery on the outcome of interest. We next explore whether those who are genetically predisposed to the outcomes are differentially affected. For example, those who are at higher genetic risk of type-2 diabetes may be more affected by prenatal sugar exposure compared to those who are genetically at lower risk. Alternatively, those at higher genetic risk may be more protected by the high-sugar environment as these fetuses may be ``programmed'' to cope better with later-life abundant nutrition. We explore this by modelling the gene-environment interplay in our empirical specification as:
\begin{equation}
\label{eq:GxE}
y_i = \beta_0 + \beta_1 G_i + \beta_2 E_i + \beta_3 G_i \times E_i + \gamma X_i + f_1(X_i,E_i) + f_2(X_i,G_i) + \sum_{p=1}^{10}\delta_p PC_i^p + \varepsilon_i
\end{equation}

\noindent
where $G_i$ denotes the polygenic score that is specific to outcome $y_i$. The right-hand side also includes the first 10 principal components of the genetic data to account for ancestry (Price et al., 2006).\footnote{Recall that \hyperref[sec:AppendixB]{\ref*{sec:AppendixB}} includes a brief discussion of some of the genetic concepts and jargon and explains in more detail how the polygenic scores are created. The results in the paper are strongly robust to the exclusion of the principal components.} Furthermore, next to the interactions $f_1(X_i,E_i)$ between covariates $X_i$ and the exposure indicator $E_i$, we add interactions $f_2(X_i,G_i)$ between the covariates $X_i$ and the polygenic score $G_i$. This prevents that effects of such interactions are inadvertently picked up by the estimated effects of the $G_i \times E_i$ interaction (see Keller, 2014).\footnote{For reasons discussed above, we do not add the interaction between the month effects and $E_i$, but we do include the interactions between the month effects and the polygenic score $G_i$.} 

Note that the above specification resembles a health production function with gene-environment interactions. The coefficients of prime interest are $\beta_2$ and $\beta_3$. The parameter $\beta_2$ replaces the parameter $\alpha_1$ in the baseline specification. The parameter $\beta_3$ captures whether those who are genetically at a higher risk of the outcome are more (or less) likely to be affected by the exposure, as compared to those at a lower genetic risk.\footnote{As we apply an ITT design and we do not observe actual individual consumption levels early in life, one needs to be careful with the interpretation of interaction effects. Specifically, for the interaction effect of $G_i$ and actual consumption to have the same sign as the interaction effect between $G_i$ and the exposure instrument $E_i$, the genetic background should not strongly moderate the effect of exposure on consumption. This is likely to be satisfied e.g. if actual consumption after derationing remains restricted by a limited availability of affordable confectionery, as is the case in our setting. Either way, one might argue that what matters from a policy evaluation perspective are variables that can be directly manipulated, which in our case are the environment indicators (i.e., the rationing of confectionery) rather than individuals' actual confectionery choices.} The parameter $\beta_1$ captures the change in $y_i$ for a one standard deviation change in $G_i$.\footnote{Note that $G_i$ may also capture parental genetic predispositions that can shape the child's environment (so-called ``genetic nurture''; see e.g. Bates et al., 2018, and Kong et al., 2018; see also Biroli et al., 2022, for a detailed discussion and interpretation of $G \times E$ studies). In that case, 
the estimate of $\beta_1$ may potentially reflect effects of certain environmental characteristics that are correlated with genetic endowments. Since genetic variants are randomly distributed across siblings, one way of isolating the direct (causal) genetic effect is by way of a within-family analysis. Although in some study settings this has been possible in the UK Biobank -- there are approximately 40,000 siblings -- it is too restrictive in our setting, since it relies on observing families where one sibling is exposed to the derationing and the other is not.}


\section{Results}
\label{sec:Results}

\autoref{tab:reg1} presents the baseline results for the outcome variables. Panel A presents the estimates from regressions that exclude area fixed effects, whilst panels B, C and D report the results from regressions that include county (of which there are 65), administrative county (232), and district of birth (1434) fixed effects, respectively. The results in Panel A suggest that in general the derationing (i.e., more confectionery consumption) leads to better outcomes along some dimensions while it does not affect the other outcomes. Importantly, being exposed to derationing has a positive effect on years of education. More specifically, being in utero during the derationing led to an increase in years of education of 0.17, so about 2 months on an annual base. Note that this result controls for the season of birth. We also find that BMI is 0.09 units lower and sugar intake is 1\% or 0.2 percentage points less for those exposed to the derationing compared to those not exposed.\footnote{A 0.09 drop in BMI is similar to a 0.25kg drop in weight for an individual of average (1.68m) height in the data.} 

We find no statistically significant effects on height, cardiovascular disease, type-2 diabetes or on carbohydrate or fat intake. Indeed, most of these coefficients are relatively precisely estimated at values close to zero. Regarding birth weight we find a 14g increase for those exposed to confectionery derationing, though this is not statistically significant in Panel A. Controlling for county (Panel B), administrative county (Panel C) or district of birth (Panel D) fixed effects. does not affect any of these findings in a notable way.

\begin{center}
	\textbf{[\autoref{tab:reg1} about here]}
\end{center}

The results on BMI and sugar intake, and to some extent those on education and birth weight, are in line with the ``fetal programming'' hypothesis or the more general ``developmental origins of health and disease'' framework to explain later-life health outcomes. This stipulates that late-life health may benefit from a pre- and neonatal environment that is well aligned to the later-life environment. In our setting, a high-sugar environment in utero is better aligned with abundant nutrition later in
life. Such a high-sugar environment in utero then ``programs'' the fetus to better cope with abundant-nutrition environments later in life, thereby improving its later-life outcomes. 

Essentially the same argument can be made when considering confectionery intake during pregnancy as a means to reduce stress. In the absence of confectionery food, cravings during pregnancy may not be met, which may increase stress levels of the mother and thereby prepare the fetus for a life full of stress. This could decrease birth weight and increase BMI and sugar intake later in life.\footnote{Recent research on mice has shown that sugar intake of adults increases the activation of tissue-resident memory T cells in the intestinal wall in order to fight infections. This leads to faster clearance of an infection (Konjar et al., 2022). Among pregnant women this mechanism may explain cravings for sweet food. Either way, according to the developmental origins framework, a lower disease occurrence during gestation protects the offspring against adverse health outcomes later in life. Such a mechanism may amplify the stress-related pathway mentioned in the main text.}

A shortcoming of explanations based on fetal programming is that height, cardiovascular disease and type-2 diabetes are found to not respond to derationing. These three outcomes are the outcomes that have been found to be most responsive to early-life conditions in the literature. Of course, the exposure we consider differs somewhat from the in utero exposures in the literature, such as famines, stressful bereavements, recessions, epidemics and seasonal variation in food availability. The fact that we do find  
effects on later-life sugar intake suggests that adult food preferences depend on the in-utero environment. This may in turn be a mediator of effects on other outcomes, in a way that differs from the mechanisms behind effects of the more typical early-life exposures.

A slightly different explanation for our findings is that access to confectionery boosted pregnant women's happiness, which in turn could have led to an improvement in bonding with the newborn and hence more favorable educational and late-life health outcomes of the offspring.\footnote{Indeed, the consumption of sugar has been associated with reduced stress-induced cortisol (Tryon et al., 2015), and prenatal stress has been linked to an increase in adverse child outcomes (Gitau et al., 1998; Aizer, Stroud and Buka, 2016; Persson and Rossin-Slater, 2018). Therefore, a higher consumption of sweets could have provided the mother with psychological benefits with subsequent effects on their children.} In this case the nutritional value of confectionery is not of prime relevance.

The coefficient estimates in \autoref{tab:reg1} reflect average effects for our analysis sample. To proceed, we explore whether those who are genetically predisposed to a trait are differentially affected by the environmental prenatal exposure, by 
incorporating the gene-environment interplay in the empirical model. Panels A and B of \autoref{tab:reg2} present the estimates without and with administrative county fixed effects, respectively. 
The additive effects of prenatal exposure to derationing are very similar to those shown in \autoref{tab:reg1} above. Also, not surprisingly, the additive effects of the respective polygenic scores are strongly positive on all outcomes of interest, and the results are robust to including area fixed effects. A one standard deviation increase in the polygenic score for educational attainment, for example, is associated with an increase of 1.2-1.3 years of schooling; consistent with the existing literature (see e.g. Muslimova et al., 2020). Similarly, a one standard deviation increase in the polygenic score for birth weight increases birth weight by $\approx$100 grams, and a one standard deviation increase in the polygenic score for BMI leads to an increase of 1.5 units. For height, the estimates show an effect of $\approx$3.4 cm for each standard deviation increase in the polygenic score, and for cardiovascular disease and type-2 diabetes, we find increases of five and two percentage points respectively. Finally, we find that a one standard deviation increase in the polygenic score for the macronutrients increases the consumption of sugar, carbohydrates and fats (as a proportion of the total energy intake) by 0.4, 0.5, and 0.3 percentage points, respectively. 

More importantly, we find evidence of $G \times E$ effects. Consider the sugar density in older age for individuals with an average polygenic score (i.e., of 0). This density is 0.2 percentage points lower among those prenatally exposed to derationing compared to those not exposed. \autoref{tab:reg2} shows that this decrease is larger than 0.2 for those with a higher polygenic score for sugar. In other words, while a one standard deviation increase in the polygenic score for sugar consumption increases individuals' intake by 0.4 percentage points, we find that this effect is approximately 50\% (i.e., 0.2 percentage points) smaller among those exposed to prenatal derationing. 
Thus, a high sugar intake in utero causes people with a genetic preference for sugar to dampen their later-life consumption of sugar. In a fetal-programming perspective, it seems that a favorable in-utero environment is particularly protective for individuals who have a high genetic propensity to consume unhealthy food later in life. This makes sense from an evolutionary point of view. Put bluntly, individuals who by their genetic constitution do not consume sugar as adults do not need to be prepared by certain early-life exposures that are aimed at influencing sugar consumption. Note that we find a significant negative $G \times E$ interaction effect for other outcomes as well, including carbohydrate intake and cardiovascular disease (the latter only in Panel B). This again suggests that although a high polygenic score increases one's carbohydrate intake and risk of disease, this risk is reduced for those programmed to better survive in high-sugar environments. 

For the other outcomes, we find no heterogeneous response of exposure to derationing with respect to one's genetic predisposition for the trait. In other words, there is no evidence that one's genetic endowment for education, birth weight, BMI, height, type-2 diabetes and fat intake aggravates or alleviates individuals against a high sugar prenatal environment. Indeed, the interaction effects for these specifications are close to zero and relatively precisely estimated.\footnote{Polygenic scores are measured with error. One way of dealing with this measurement error is by using an instrumental variables (IV) specification, using polygenic scores obtained from GWAS summary statistics based on external (i.e., non-UKB) data (DiPrete et al., 2018; van Kippersluis et al., 2021). External GWAS summary statistics are available for all our outcomes other than the macronutrient intakes (Okbay et al., 2016; Locke et al. 2015; Morris et al. 2012; Wood et al., 2014; Schunkert et al., 2014). We therefore create two sets of polygenic scores for each outcome (other than macronutrients) and use Obviously-Related Instrumental Variables (ORIV; Gillen et al., 2019), specifying each polygenic score as an instrument for the other. As expected, this increases the predictive power of the polygenic score and typically increases the coefficient on the interaction term, but also its standard error. Hence, it does not lead to substantially different conclusions with respect to the $E$ or $G \times E$ effects (results not shown here but available upon request).}

\begin{center}
	\textbf{[\autoref{tab:reg2} about here]}
\end{center}


\section{Robustness analysis}\label{sec:robustness}

The following subsections discuss a series of additional analyses and sensitivity checks to ensure that our analysis is robust to different assumptions. This includes an analysis of trimester-specific effects, sensitivity to the timing of the derationing period, to accounting for potential selection into pregnancy, and to alternative specifications of the calender-time trend covariate. We also explore possible gender differences in the treatment effect. For each of these, we focus on the main effect of exposure to the derationing, rather than the $G \times E$ effects (the latter are available upon request). All analyses control for administrative county fixed effects, unless otherwise stated.

\subsection{Trimester-specific exposure}

Our baseline analysis above estimates effects of being exposed to confectionery derationing averaged over all those in utero during this period. However, the literature on the long-term effects of intrauterine exposures highlights potentially heterogeneous effects due to exposure at different stages of the pregnancy, generally distinguishing between three trimesters (see e.g. Lumey, Stein and Susser, 2011; Almond and Mazumder, 2011). For example, most organs are formed during the first trimester, so adverse circumstances in this period may primarily affect organ development. 

To explore whether there are differential effects of exposure by trimester, we extend the baseline specification by allowing for three separate ``trimester effects''. This requires a mapping from the actual pregnancy interval into the gestational trimester $j$ of exposure. In this we follow the literature on the effects of the Dutch Hunger Winter famine which also lasted about 4 months (Lumey, Stein and Susser, 2011). In equation (\ref{eq:E}), we replace $\alpha_1 E_i$ by $\sum_{j=1}^3 \alpha _{j} E_{ij}$, where each $E_{ij}$ is binary and $\sum_{j=1}^3  E_{ij} \leq 1$.\footnote{Exposure in the first trimester is defined as one for those born between December 24, 1949 and May 12, 1950, and zero otherwise. Those born between September 24, 1949 and December 23, 1949 are defined as being exposed in the second trimester, and those born between April 24, 1949 and September 23, 1949 are defined as being exposed in the third trimester.} The results are presented in Panel A of \autoref{tab:reg3}. This shows that the positive effect on years of education is largely driven by exposure in the third trimester, with this leading to an increase of 0.2 years of schooling, with a negative (insignificant) effect for first trimester exposure. The (marginally significant) positive effect on birth weight is driven by exposure in the first and second trimester. Finally, the negative effect on BMI is mainly driven by exposure during the second and third trimester, and the negative effect on sugar consumption is similar across the three trimesters. 

During the third trimester, the fetus gains most weight, so malnutrition in this trimester may cause the caloric needs to support intrauterine growth not to be fulfilled. Conditions in the UK in 1949 cannot be accurately described as malnutrition. However, appetite may depend on the type of food. Put simply, in the third trimester it may be more convenient to consume confectionery than to eat a huge plate of cooked potatoes. In settings where caloric supply is sufficient, such considerations may be relevant and may have long-run implications, which may explain some of the trimester-specific results.

\subsection{Timing of confectionery derationing}

Although confectionery derationing started on April 24, 1949 and was abolished on August 13, 1949, its effects on actual consumption may have been restricted to a narrower time interval. As discussed earlier in the paper, the demand for confectionery was so high that there were long queues outside shops, with evidence that, by May 1949, some confectionery shops had run out of stock. This would imply that some individuals who are currently defined as treated are in fact not exposed to confectionery prenatally due to a non-availability, potentially attenuating our estimate of interest. To explore this in more detail, we re-run the analysis, setting $E_i$ equal to one for individuals who were in utero between April 24, 1949 and the end of \textit{May 1949} (instead of August 13, 1949). 
The estimates are presented in Panel B of \autoref{tab:reg3}. These confirm the positive effect of derationing exposure on years of education and birth weight. Furthermore, we again find that prenatal exposure to derationing led to a drop in BMI of 0.08 units (equivalent to a reduction in body weight of 0.23 kg for a person of average height) and a 0.2 percentage point drop in the sugar density of the diet. All these results are consistent with those found for the baseline specification.

In Panel C, we additionally report the trimester-specific effects when exposure is defined as having been in utero between April 24, 1949 and the end of May 1949.\footnote{First trimester exposure is now defined as those born between November 13, 1949 and February 28, 1950. Those born between August 13, 1949 and November 12, 1949 are defined as being exposed in the second trimester, and those born between April 24, 1949 and August 12, 1949 are defined as being exposed in the third trimester.} The findings are broadly consistent with those above, with second trimester exposure for birth weight being most important, third trimester exposure for BMI, and first trimester exposure for height. We also find a significant effect for years of education of second trimester exposure, which is also driving the effect on sugar intake.

\begin{center}
	\textbf{[\autoref{tab:reg3} about here]}
\end{center}

\subsection{Timing of pregnancy}
We next explore the assumption that pregnancies are not timed relative to the derationing period. 
For this we exclude individuals who (we infer) were not yet conceived at the time of the government \textit{announcement} (as opposed to \textit{implementation}) of confectionery derationing on 21 February 1949 but who were conceived before the end of derationing. In effect, we compare pregnancies which were unexposed to derationing (including all those born before April 24, 1949) to those who were already pregnant when the government announced the policy, removing any potential selection into pregnancy due to the announced derationing. The latter group now includes those born between April 24, 1949 and November 24, 1949 (i.e., 9 months after the announcement on 21 February 1949). \autoref{tab:timing} presents the estimates, showing a positive effect on education and a negative effect on BMI and sugar intake. The magnitude of the effects is very similar to those shown above, suggesting there is no selection into pregnancy after the government announcement to deration confectionery.

\begin{center}
	\textbf{[\autoref{tab:timing} about here]}
\end{center}

\subsection{Alternative functional forms for calendar-time trend effects}
With a relatively small observation window (those born between April 24, 1947 and May 24, 1952), most outcomes do not display strong non-linear trends in birth date among the non-exposed, suggesting that linear specifications in the year-month of birth are sufficient. We test the robustness of this in \autoref{tab:difftrend}. Here, Panel A only accounts for a linear function in the year-month of birth (i.e., not allowing this to differ by exposure status). In Panel B a quadratic function in the year-month of birth is assumed whereas in Panel C we allow for differential quadratic functions by exposure status. The results are generally consistent to those above: being exposed to the derationing in utero increases individuals' years of schooling, with some evidence that it reduces BMI and sugar consumption later in life. Although the magnitude of the point estimates are similar to those above, they are not always statistically significantly different from zero. The exception is birth weight, for which the estimates are less robust to the different functional form specifications. 

\begin{center}
	\textbf{[\autoref{tab:difftrend} about here]}
\end{center}

\subsection{Heterogeneous effects}
Finally, we explore heterogeneity of the treatment effects with respect to gender. Male fetuses are known to be more vulnerable than female fetuses, with the latter being more likely to survive adverse intrauterine conditions (see e.g., Kraemer, 2000). As the derationing is likely to have led to a relatively mild change in the prenatal environment, it is unlikely to cause differential survival between the genders. \autoref{tab:reg4} reports estimates separately for women (Panel A) and men (Panel B). This shows that although both genders show a positive effect on years of education, it is larger and only significant for men, whilst the positive effect on birth weight is driven by women. We find a reduction in BMI for both genders, though with the reduced sample sizes and inflated standard errors they are not significantly different from zero at conventional levels. Finally, although the negative effect on the share of energy from sugar is found for both genders, it is larger and significantly different from zero only for men. 

\begin{center}
	\textbf{[\autoref{tab:reg4} about here]}
\end{center}


\section{Conclusions}

Due to high rates of obesity and diet-related health problems in the developed world, the potential consequences of excessive sugar intake have received much attention in academic research, with governments across the world trying to encourage individuals to reduce their sugar consumption. Indeed, the WHO recommends individuals to reduce their daily sugar intake to under 10\% of their total energy intake and it claims that a further reduction to a level below 5\% would provide additional health benefits (WHO, 2015). These guidelines mainly refer to the \textit{short term} health benefits of reduced sugar intake. The \textit{longer term} effects on health and well-being are much less known. In particular, there is no evidence on the long term causal effects of sugar exposure in utero. 

Our analysis has directly targeted this omission in the literature. We exploit a temporary increase in the availability of confectionery in a period that is characterized by a mostly time-invariant availability of food items, where the nutritional composition of diets remained fixed due to the majority of foods being rationed. 
We find that the prenatal exposure to confectionery derationing increased education by $\approx$0.15 years (1.8 months), lowered later-life BMI by 0.08 units and reduced later-life sugar consumption by 0.2 percentage points (0.9\%), with some evidence that it also led to an increase in birth weights of approximately 10-15 grams. 

The results on BMI and sugar intake, and to some extent those on education and birth weight, are in line with the ``fetal programming'' or ``developmental origins of health and disease'' framework to explain later-life health outcomes. Along these lines, a high-sugar environment in utero ``programs'' the fetus to better cope with abundant-nutrition environments later in life, thereby improving its later-life outcomes. Essentially the same argument can be made when considering confectionery intake during pregnancy as a means to reduce stress. In the absence of confectionery food, cravings during pregnancy may not be met, which may increase stress levels of the mother and thereby prepare the fetus for a more stressful life. 

The programming explanation is more difficult to reconcile with the absence of effects on other outcome variables. Here it should be kept in mind that, in contrast to the literature on long-run effects of early-life conditions, we are not analyzing omnibus measures of adversity (such as famines or recessions or the availability of common staple foods) but we focus on access to a highly specific food type. The fact that we find effects on later-life sugar intake suggests that adult food preferences depend on the in-utero environment. This may in turn be a mediator of effects on other outcomes. We view it as an important topic for future research to examine long-run effects of a wide range of specific food types in order to enhance our understanding of their roles in effects on later-life outcomes. 
An advantage of our setting is that individuals were unlikely to substitute or complement the increase in confectionery with changes in other food intake, since most other foods were rationed. This implies that we identify the effects of increased prenatal exposure to confectionery \textit{per se}, without the potential simultaneous adjustments to other food items (though it is possible that our estimates partially capture a reduced exposure to artificial sweeteners such as saccharin). 

We also explored whether there is evidence that one's genetic predisposition can exacerbate or alleviate the effects. To do so, we use the molecular genetic data available in the UK Biobank, constructing polygenic scores for each outcome of interest and interacting these with the treatment effect. We find some evidence of genetic heterogeneity of treatment effects. More specifically, with high polygenic scores for cardiovascular disease, sugar and carbohydrate intake in a high-sugar prenatal environment, the beneficial effects of derationing are higher. In a fetal-programming perspective, it seems that a favorable in-utero environment is particularly protective for individuals who have a high genetic propensity for adverse late-life outcomes. This may make sense from an evolutionary point of view: individuals who by their genetic constitution face favorable late-life outcomes may not need to be do programmed through certain early-life exposures.


\newpage

\addcontentsline{toc}{section}{References}

\section*{References}

\newlength{\leftlocal}
\setlength{\leftlocal}{\leftmargini}
\addtolength{\leftmargini}{-.5\leftmargini}

\begin{description}

\newlength{\labellocal}
\setlength{\labellocal}{\labelwidth}
\setlength{\labelwidth}{5pt}

\newlength{\itemlocal}
\setlength{\itemlocal}{\itemsep}
\setlength{\itemsep}{0pt}

\item Aizer, A., L. Stroud, and S. Buka. (2016), Maternal stress and child outcomes: Evidence from
siblings. {\em Journal of Human Resources}, 51(3), 523--555.

\item Almond, D. and Currie, J. (2011), Killing me softly: The fetal origins hypothesis, {\em Journal of Economic Perspectives} 25, 153--172.

\item Almond, D. and B. Mazumder, (2011), Health capital and the prenatal environment: The effect of Ramadan observance during pregnancy, {\em American Economic Journal: Applied Economics} 3, 56--85.

\item Avery, A.R. and Duncan, G.E. (2019),  Heritability of Type 2 Diabetes in the Washington State Twin Registry, {\em Twin Research and Human Genetics} 22.2, 95--98.

\item Barcellos, H.S., Carvalho, L.S. and Turley, P. (2018), Education can reduce health differences related to genetic risk of obesity,  {\em Proceedings of the National Academy of Sciences} 115.42, E9765--E9772.

\item Bates, T. C., Maher, B. S., Medland, S. E., ..., Gillespie, N.A. (2018), The nature of nurture: using a virtual-parent design to test parenting effects on children's educational attainment in genotyped families, {\em Twin Research and Human Genetics} 21.2, 73--83.

\item B\'edard, A., Northstone, K., Henderson, J. and Shaheen, S.O. (2017), Maternal intake of sugar during pregnancy and childhood respiratory and atopic outcomes, {\em European Respiratory Journal} 50: 1700073.

\item Bengtsson, T. and Lindstr\"om, M. (2000), Childhood misery and disease in later life: The effects on mortality in old age of hazards experienced in early life, southern Sweden, 1760--1894, {\em Population Studies} 54, 263--277.

\item Bengtsson, T. and Lindstr\"om, M. (2003), Airborne infectious diseases during infancy and mortality in later life in southern Sweden, 1766--1894, {\em International Journal of Epidemiology} 32, 286--294.

\item van den Berg, G.J. and Modin, B. (2013), Economic conditions at birth, birth weight, ability, and the causal path to cardiovascular mortality, Working paper, IZA Bonn.

\item van den Berg, G.J., Pinger, P.R. and Schoch, J. (2016), Instrumental variable estimation of the causal effect of hunger early in life on health later in life, {\em Economic Journal}, 126, 465--506.

\item Biroli, P., T.J. Galama, S. von Hinke, H. van Kippersluis, C.A. Rietveld, and K. Thom (2022), The economics and econometrics of gene-environment interplay, {\em arXiv: 2203.00729}, 
\url{https://doi.org/10.48550/arXiv.2203.00729}

\item Biroli, P. and Z\"und, C. (2022), Genes, pubs and drinks: Gene-environment interplay and alcohol licensing policy in the United Kingdom. {\em Mimeo, University of Bologna}

\item Bocarsly, M.E., Barson, J.R., Hauca, J.M., Hoebel, B.G., Leibowitz, S.F. and Avena, N.M. (2012), Effects of perinatal exposure to palatable diets on body weight and sensitivity to drugs of abuse in rats, {\em Physiological behavior} 107, 568--575.

\item Branigen, A.R., McCallum, K. J. and Freese, J. (2013), Variation in the heritability of educational attainment: an international meta-analysis, {\em Social Forces} 92, 109--140.

\item Burnett, J. (1987), Plenty and Want: A social history of diet in England from 1815 to the present day, Routledge.

\item Casas, R., Castro Barquero, S. and Estruch, R. (2020), Impact of sugary food consumption on pregnancy: a review, {\em Nutrients} 12:3574. 

\item Cohen, J., Rifas-Shiman, S.L., Young, J., and Oken, E. (2018), Associations of prenatal and child sugar intake with child cognition, {\em American Journal of Preventive Medicine} 54, 727--735. 

\item Collins, R., Peakman, T., Alegre-Diaz, J., ..., (2012),  What makes UK Biobank special?, {\em Lancet} 379, 1173--1174 

\item Davies, N. (2014), Europe: A history, Random House.

\item Department of Health (2002), Scientific Review of the Welfare Food Scheme, {\em Report on Health and Social Subjects} 51.

\item DiPrete, T.A., Burik, C.A.P. and Koellinger, P.D. (2018), Genetic instrumental variable regression: Explaining socioeconomic and health outcomes in nonexperimental data. {\em Proceedings of the National Academy of Sciences} 115, E4970--E4979.

\item Domingue, B.W., Trejo, S., Armstrong-Carter, E., and Tucker-Drob, E.M. (2020), Interactions between polygenic scores and environments: Methodological and conceptual challenges, {\em Sociological Science} 7, 465--486.

\item Fletcher, J.M. (2012). Why have tobacco control policies stalled? Using genetic moderation to examine policy impacts. {\em PLOS One} 7(12): e50576.

\item Fletcher, J.M. (2019), Environmental bottlenecks in children's genetic potential for adult socio-economic attainments: Evidence from a health shock, {\em Population Studies} 73, 139--148.

\item Financial Times (1949), Changes in the confectionery industry: Complete derationing urged by manufacturers, Tuesday February 15, 1949. 

\item Financial Times (1949), All sweets freed, Tuesday February 22, 1949. 

\item Financial Times (1949), Retail trade up in April, Saturday June 11, 1949. 

\item Financial Times (1949), Sweets and de-rationing problems, Thursday June 16, 1949. 

\item Fitzgerald, R. (1995), Rowntree and the marketing revolution, 1862-1969, Cambridge University Press.

\item Geissler, C. and Oddy, D.J. (1993), Food, diet and economic change: Past and present, Leicester University Press.

\item Gillen, B., Snowberg, E., and Yariv, L. (2019), Experimenting with measurement error: Techniques with applications to the Caltech Cohort Study. {\em Journal of Political Economy}, 127(4), 1826--1863.

\item Gitau, R., Cameron, A., Fisk, N. M. and Glover, V. (1998), Fetal exposure to maternal cortisol. {\em The Lancet}, 352, 707--708.

\item Goldring, S., Henretty, N., Mills, J., Johnson, K. and Smallwood, S. (2011), Mortality of the ``Golden Generation'': What can the ONS Longitudinal Study tell us?, 
{\em Population Trends} 145, 203--232.

\item Goran, M., Plows, J. and Ventura, E. (2018), Effects of consuming sugars and alternative sweeteners during pregnancy on maternal and child health: Evidence for a secondhand sugar effect, {\em Proceedings of the Nutrition Society} 78, 1--10. 

\item von Hinke, S. and S{\o}rensen, E. (2022), The long-term effects of early-life pollution exposure: Evidence from the London Smog, {\em arXiv:2202.11785}

\item Horikoshi, M., Beaumont, R., Day, F., ..., Freathy, R. (2016), Genome-wide associations for birth weight and correlations with adult disease, {em\ Nature}, 538, 248--252.

\item Keeton, R.B. (2011), Price Ceilings and Rationing: The Base Ingredients of the Black Market Food Industry in Nevada During World War II, {\em Psi Sigma Siren} 7.1, 6. 

\item Keller, M. (2014), Gene $\times$ environment interaction studies have not properly controlled for potential confounders: the problem and the (simple) solution, {\em Biological Psychiatry}, 75(1): 18-24.

\item van Kippersluis, H. et al., (2021) Stop Meta-Analyzing, Start Instrumenting: Maximizing the predictive power of polygenic scores, BioRxiv, doi.org/10.1101/2021.04.09.439157.

\item Kong, A., Thorleifsson, G., Frigge, M., ..., Stefansson, K. (2018), The nature of nurture: Effects of parental genotypes, {\em Science} 359, 424--428.

\item Konjar, S., Ferreira, C., Carvalho, F.S., ..., Veldhoen, M. (2022), Intestinal tissue-resident T cell activation depends on metabolite availability, {\em Proceedings of the National Academy of Sciences}, 119(34): e2202144119, 1--12.

\item Kraemer, S. (2000) The fragile male. {\em British Medical Journal}, 321(7276): 1609--1612.

\item Lee, J.J., Wedow, R., Okbay, A., ..., Cesarini, D. (2018), Gene discovery and polygenic prediction from a 1.1-million-person GWAS of educational attainment. {\em Nature Genetics}, 50(8): 2212--1121.

\item Lindeboom, M., Portrait, F.R.M. and van den Berg, G.J. (2010), Long-run effects on longevity of a nutritional shock early in life: the Dutch Potato Famine of 1846-1847, {\em Journal of Health Economics} 29, 617--629.

\item Locke, A., B. Kahali, S. Berndt, ..., E.K. Speliotes, (2015), Genetic studies of body mass index yield new insights for obesity biology, {\em Nature} 518, 197-–206.

\item Lucchini, M., Della Bella, S. and Pisati M. (2013), The weight of the genetic and environmental dimensions in the inter-generational transmission of educational success, {\em European Sociological Review} 29.2, 289--301.

\item Lumey, L.H., Stein, A.D., and Susser, E. (2011), Prenatal famine and adult health, {\em Annual Review of Public Health}, 32: 24.1--24.26.

\item Maes, H. H., Neale, M. C., and Eaves, L. J. (1997), Genetic and environmental factors in relative body weight and human adiposity, {\em Behavior Genetics} 27, 325--351. 

\item MAF 156/262 (1947), Rationing statistics: Straight rationing (Part I), 1946-1947. The National Archives.

\item MAF 156/263 (1951), Rationing statistics: Straight rationing (Part II), 1948-1951. The National Archives.

\item MAF 156/270 (1952), WWII National Food Survey, 1952. Cabinet and general monthly report: Points rationing and welfare foods, 1947-1952, The National Archives.

\item Marenberg, M.E., Risch, N., Berkman, L.F., Floderus, B. and de Faire, U. (1994), Genetic susceptibility to death from coronary heart disease in a study of twins, {\em The New England journal of medicine} 330, 1041--1046.

\item Maruyama, S. and E. Heinesen (2020), Another look at returns to birthweight, {\em Journal of Health Economics} 70, 102269.

\item Ministry of Food (1951), The urban working-class household diet 1940-1949: First report of the National Food Survey Committee. London, His Majesty's Stationery Office.

\item Ministry of Health (1946), On the state of the public health during six years of war. London, His Majesty's Stationery Office.

\item Muslimova, D., H. van Kippersluis, C.A. Rietveld, S. von Hinke, and S.F.W. Meddens, (2020), Dynamic complementarity in skill production: Evidence from genetic endowments and birth order, {\em arXiv:2012.05021} 

\item Morris, A.P., B.F. Voight, T.M. Teslovich, ..., and M.I. McCarthy (2012), Large-scale association analysis provides insights into the genetic architecture and pathophysiology of type 2 diabetes,{\em Nature Genetics}, 44(9), 981--990.

\item Okbay, A., Beauchamp, J. P., Fontana, M. A., ..., Salomaa, V. (2016), Genome-wide association study identifies 74 loci associated with educational attainment, {\em Nature} 533.7604, 539--542.

\item Okbay, A., Wu, Y., ..., Young, A. (2022), Polygenic prediction of educational attainment within and between families from genome-wide association analyses in 3 million individuals. {\em Nature Genetics} 54, 437--449

\item Palatnik, A., Moosreiner, A. and Olivier-Van Stichelen, S. (2020), Consumption of non-nutritive sweeteners during pregnancy, {\em American Journal of Obstetrics and Gynecology}, 223, 211--218.

\item Pereira, R., C.A. Rietveld, and H. van Kippersluis (\textit{in press}), The interplay between maternal smoking and genes in offspring birth weight, {\em Journal of Human Resources}

\item Phelan, S., Hart, C., Phipps, M., Abrams, B., Schaffner, A., Adams, A. and Wing, R. (2011), Maternal behaviors during pregnancy impact offspring obesity risk, {\em Experimental Diabetes Research}, 2011: 985139. 

\item Plomin, R. and Bergeman, C. S. (1991), The nature of nurture: Genetic influence on ``environmental'' measures, {\em Behavioural and Brain Sciences} 14, 414--427.

\item Price, A. L., Patterson, N. J., Plenge, R. M., ..., Reich, D. (2006), Principal components analysis corrects for stratification in genome-wide association studies, {\em Nature genetics} 38.8, 904--909.

\item Rietveld, C. A., Medland, S. E., Derringer, ...,  Preisig, M. (2013), GWAS of 126,559 individuals identifies genetic variants associated with educational attainment, {\em Science} 340.6139, 1467--1471.

\item Rivett, G. (2014), {\em From Cradle to Grave: The history of the NHS 1948-1987}, Nuffield Trust, London.

\item Roseboom, T., de Rooij, S., Painter, R. (2006), The Dutch famine and its long-term consequences for adult health, {\em Early Human Development}, 82(8): 485-491.

\item Royle, D. (2020), Sweet rationing in World War Two, Working paper, Sawbridgeworth Local History Society.

\item Schmitz, L. and D. Conley, (2016), The long-term effects of Vietnam-era conscription and genotype on smoking behavior and health. {\em Behavior Genetics}, 46(1), 43--58.

\item Scholte, R.S., G.J. van den Berg, M. Lindeboom (2015), Long-run effects of gestation during the Dutch hunger winter famine on labor market and hospitalization outcomes. {\em Journal of Health Economics}, 39(C): 17--30.

\item Schulz, L.C. (2010), The Dutch Hunger Winter and the developmental origins of health and disease, {\em PNAS} 107, 16757--16758.

\item Sudlow, C., Gallacher, J., Allen, ..., Collins, R. (2015), UK biobank: an open access resource for identifying the causes of a wide range of complex diseases of middle and old age, {\em Plos Medicine} 12.3, e1001779.

\item Thomson, M. (2018), The birth pains of the NHS: how Britain's National Health Service was created, {\em History Magazine}, University of Warwick.

\item Tryon et al. (2015), Excessive sugar consumption may be a difficult habit to break: A view from the brain and body, {\em The Journal of Clinical Endocrinology and Metabolism}, 100(6): 2239-2247.

\item Tyrrell, J.S., Yaghootkar, H., Freathy, R.M., ..., Frayling, T.M. (2013), Parental diabetes and birthweight in 236 030 individuals in the UK biobank study., {\em International journal of epidemiology} 42.6, 1714--1723.

\item Vilhjálmsson, B.J., Yang, J., Finucane, H.K, ..., Won, H. (2015), Modeling linkage disequilibrium increases accuracy of polygenic risk scores, {\em The American Journal of Human Genetics} 97, 576--592.

\item Walsh, J.M., Mahony, R., Byrne, J., Foley, M. and McAuliffe, F.M. (2011), The association of maternal and fetal glucose homeostasis with fetal adiposity and birthweight, {\em European Journal of Obstetrics, Gynecology and Reproductive Biology} 159, 338--341.

\item WHO (2015), {\em Guideline: sugars intake for adults and children}, World Health Organization, Geneva.

\item Willet, W.C., Howe, G.R., Kushi, L. (1997), Adjustment for total energy intake in epidemiologic studies, {\em American Journal of Clinical Nutrition}, 65(suppl): 1220S-1228S.

\item Wood, A., Esko, T., Yang, J. ... T. Frayling (2014), Defining the role of common variation in the genomic and biological architecture of adult human height. {\em Nature Genetics} 46, 1173--1186 

\item Zweiniger-Bargielowska, I. (2000), Austerity in Britain: Rationing, Controls, and Consumption, 1939-1955, Oxford University Press.

\end{description}

\setlength{\leftmargini}{\leftlocal} \setlength{\labelwidth}{\labellocal} %
\setlength{\itemsep}{\itemlocal}

\newpage

\section*{Tables and Figures}

\bigskip
\begin{table}[H]
\caption{Descriptives}
\centering
{\scriptsize
\begin{tabular}{lcccccccccccccccccccc}
\toprule
			                &     N 	& 	mean &      sd  \\
\midrule
Years of education 			&    83,647 &  13.94 &  (5.140) \\
Birth weight (in g)			&    47,476 & 3331.8 &  (672.2) \\
BMI	       	  				&    83,647 &  27.57 &  (4.790) \\
Height (in cm) 				&    83,647 & 168.52 &  (9.216) \\
Cardiovascular disease		&    83,647 &   0.33 &  (0.471) \\
Type-2 diabetes 			&    83,647 &   0.05 &  (0.222) \\
Sugar intake    			&    36,131 &   0.23 &  (0.069) \\
Carbohydrate intake  		&    36,131 &   0.48 &  (0.082) \\
Fat intake      			&    36,131 &   0.33 &  (0.067) \\
Male 						& 	 84,669 &   0.45 &  (0.498) \\
\bottomrule
\addlinespace[.75ex]
\end{tabular}
}
\label{tab:descr}
\caption*{\noindent\scriptsize Sample sizes, means and standard deviations of the estimation sample, including individuals born between April 1947 (i.e., two years prior to the temporary sugar derationing) and May 1952 (i.e., two years and nine months post the end of the temporary sugar derationing). }
\end{table}

\begin{landscape}
\begin{table}[H]
\caption{Effects of prenatal exposure to confectionery derationing}
\centering
{\tiny
\begin{tabular}{lcccccccccccccccccccc}
\toprule
            &\multicolumn{1}{c}{(1)}&\multicolumn{1}{c}{(2)}&\multicolumn{1}{c}{(3)}&\multicolumn{1}{c}{(4)}&\multicolumn{1}{c}{(5)}&\multicolumn{1}{c}{(6)}&\multicolumn{1}{c}{(7)}&\multicolumn{1}{c}{(8)}&\multicolumn{1}{c}{(9)}\\
            &\multicolumn{1}{c}{\shortstack{Years of\\Education}}&\multicolumn{1}{c}{\shortstack{Birth weight\\(in g)}}&\multicolumn{1}{c}{BMI}&\multicolumn{1}{c}{\shortstack{Height\\(in cm)}}&\multicolumn{1}{c}{\shortstack{Cardiovascular\\Disease}}&\multicolumn{1}{c}{\shortstack{Type II\\Diabetes}}&\multicolumn{1}{c}{\shortstack{Sugar\\intake}}&\multicolumn{1}{c}{\shortstack{Carbohydrate\\intake}}&\multicolumn{1}{c}{\shortstack{Fats\\intake}}\\
\midrule
\textbf{Panel A: No area fixed effects}&                     &                     &                     &                     &                     &                     &                     &                     &                     \\
\addlinespace
E: \textit{In utero} during derationing&       0.167\sym{***}&      13.872         &      -0.096\sym{*}  &       0.018         &      -0.004         &      -0.003         &      -0.002\sym{*}  &      -0.001         &      -0.001         \\
            &     (0.056)         &     (9.594)         &     (0.050)         &     (0.065)         &     (0.005)         &     (0.002)         &     (0.001)         &     (0.001)         &     (0.001)         \\
\midrule
\textbf{Panel B: County FE (n=65)}&                     &                     &                     &                     &                     &                     &                     &                     &                     \\
\addlinespace
E: \textit{In utero} during derationing&       0.154\sym{***}&      14.582\sym{*}  &      -0.090\sym{**} &       0.006         &      -0.004         &      -0.003         &      -0.002         &      -0.001         &      -0.001         \\
            &     (0.050)         &     (8.330)         &     (0.034)         &     (0.049)         &     (0.005)         &     (0.002)         &     (0.001)         &     (0.001)         &     (0.001)         \\
\midrule
\textbf{Panel C: Administrative county FE (n=232)}&                     &                     &                     &                     &                     &                     &                     &                     &                     \\
\addlinespace
E: \textit{In utero} during derationing&       0.147\sym{***}&      14.399         &      -0.084\sym{**} &       0.002         &      -0.004         &      -0.003         &      -0.002\sym{*}  &      -0.002         &      -0.001         \\
            &     (0.052)         &    (10.347)         &     (0.042)         &     (0.066)         &     (0.006)         &     (0.002)         &     (0.001)         &     (0.001)         &     (0.001)         \\
\midrule
\textbf{Panel D: District FE (n=1434)}&                     &                     &                     &                     &                     &                     &                     &                     &                     \\
\addlinespace
E: \textit{In utero} during derationing&       0.142\sym{***}&      13.531         &      -0.087\sym{**} &       0.005         &      -0.003         &      -0.003         &      -0.002\sym{**} &      -0.002         &      -0.001         \\
            &     (0.051)         &     (9.645)         &     (0.043)         &     (0.064)         &     (0.006)         &     (0.002)         &     (0.001)         &     (0.001)         &     (0.001)         \\
\midrule
No. of observations&       84184         &       47476         &       84554         &       84641         &       84672         &       84637         &       36131         &       36131         &       36131         \\
 \\
\bottomrule
\addlinespace[.75ex]
\end{tabular}
}
\label{tab:reg1}
\caption*{\noindent\scriptsize Estimates from Equation (\ref{eq:E}), with the treatment dummy ($E$) specified as those born between 24 April 1949 and 12 May 1950. Robust standard errors (clustered by year-month in Panel A and by geographic area in the Panels B--D) in parentheses. * $p < 0.10$, ** $p < 0.05$, *** $p < 0.01$.}
\end{table}


\bigskip
\begin{table}[H]
\caption{$G \times E$ effect of prenatal exposure to confectionery derationing}
\centering
{\tiny
\begin{tabular}{lcccccccccccccccccccc}
\toprule
            &\multicolumn{1}{c}{(1)}&\multicolumn{1}{c}{(2)}&\multicolumn{1}{c}{(3)}&\multicolumn{1}{c}{(4)}&\multicolumn{1}{c}{(5)}&\multicolumn{1}{c}{(6)}&\multicolumn{1}{c}{(7)}&\multicolumn{1}{c}{(8)}&\multicolumn{1}{c}{(9)}\\
            &\multicolumn{1}{c}{\shortstack{Years of\\Education}}&\multicolumn{1}{c}{\shortstack{Birth weight\\(in g)}}&\multicolumn{1}{c}{BMI}&\multicolumn{1}{c}{\shortstack{Height\\(in cm)}}&\multicolumn{1}{c}{\shortstack{Cardiovascular\\Disease}}&\multicolumn{1}{c}{\shortstack{Type II\\Diabetes}}&\multicolumn{1}{c}{\shortstack{Sugar\\intake}}&\multicolumn{1}{c}{\shortstack{Carbohydrate\\intake}}&\multicolumn{1}{c}{\shortstack{Fats\\intake}}\\
\midrule
\textbf{Panel A: Without area fixed effects}&                     &                     &                     &                     &                     &                     &                     &                     &                     \\
\addlinespace
E: \textit{In utero} during derationing&       0.169\sym{***}&      14.084         &      -0.083         &       0.027         &      -0.004         &      -0.003         &      -0.002\sym{*}  &      -0.001         &      -0.001         \\
            &     (0.056)         &     (9.292)         &     (0.051)         &     (0.055)         &     (0.005)         &     (0.002)         &     (0.001)         &     (0.001)         &     (0.001)         \\
\addlinespace
PGS (UKB)   &       1.305\sym{***}&     101.801\sym{***}&       1.549\sym{***}&       3.408\sym{***}&       0.049\sym{***}&       0.021\sym{***}&       0.004\sym{***}&       0.005\sym{***}&       0.003\sym{***}\\
            &     (0.017)         &     (3.449)         &     (0.018)         &     (0.021)         &     (0.002)         &     (0.001)         &     (0.000)         &     (0.000)         &     (0.000)         \\
\addlinespace
GxE         &       0.021         &       7.920         &      -0.020         &      -0.037         &      -0.007         &      -0.000         &      -0.002\sym{**} &      -0.003\sym{**} &       0.000         \\
            &     (0.042)         &     (8.172)         &     (0.032)         &     (0.042)         &     (0.004)         &     (0.002)         &     (0.001)         &     (0.001)         &     (0.001)         \\
\midrule
\textbf{Panel B: Administrative county FE}&                     &                     &                     &                     &                     &                     &                     &                     &                     \\
\addlinespace
E: \textit{In utero} during derationing&       0.154\sym{***}&      14.581         &      -0.075\sym{*}  &       0.016         &      -0.004         &      -0.003         &      -0.002\sym{*}  &      -0.002         &      -0.001         \\
            &     (0.050)         &    (10.385)         &     (0.040)         &     (0.054)         &     (0.006)         &     (0.002)         &     (0.001)         &     (0.001)         &     (0.001)         \\
\addlinespace
PGS (UKB)   &       1.196\sym{***}&     101.592\sym{***}&       1.532\sym{***}&       3.371\sym{***}&       0.048\sym{***}&       0.020\sym{***}&       0.004\sym{***}&       0.005\sym{***}&       0.003\sym{***}\\
            &     (0.024)         &     (3.191)         &     (0.019)         &     (0.017)         &     (0.002)         &     (0.001)         &     (0.000)         &     (0.001)         &     (0.000)         \\
\addlinespace
GxE         &       0.019         &       8.587         &      -0.021         &      -0.029         &      -0.006\sym{*}  &      -0.000         &      -0.002\sym{*}  &      -0.003\sym{**} &       0.000         \\
            &     (0.041)         &     (7.449)         &     (0.037)         &     (0.049)         &     (0.004)         &     (0.002)         &     (0.001)         &     (0.001)         &     (0.001)         \\
\midrule
No. of observations&       84184         &       47476         &       84554         &       84641         &       84672         &       84637         &       36131         &       36131         &       36131         \\
 \\
\bottomrule
\addlinespace[.75ex]
\end{tabular}
}
\label{tab:reg2}
\caption*{\noindent\scriptsize Estimates from Equation (\ref{eq:GxE}), with the treatment dummy ($E$) specified as those born between 24 April 1949 and 12 May 1950 and the polygenic score $G$ obtained from using the summary statistics from our tailor-made GWAS on the excluded UKB sample and any siblings. Panel A reports the analysis without controlling for area of birth fixed effects; Panel B reports the analysis that controls for administrative county of birth fixed effects. Robust standard errors (clustered by year-month in Panel A and by administrative county in Panel B) in parentheses. * $p < 0.10$, ** $p < 0.05$, *** $p < 0.01$.}
\end{table}


\begin{table}[H]
\caption{Timing of derationing and trimester-specific results}
\centering
{\tiny
\begin{tabular}{lcccccccccccccccccccc}
\toprule
            &\multicolumn{1}{c}{(1)}&\multicolumn{1}{c}{(2)}&\multicolumn{1}{c}{(3)}&\multicolumn{1}{c}{(4)}&\multicolumn{1}{c}{(5)}&\multicolumn{1}{c}{(6)}&\multicolumn{1}{c}{(7)}&\multicolumn{1}{c}{(8)}&\multicolumn{1}{c}{(9)}\\
            &\multicolumn{1}{c}{\shortstack{Years of\\Education}}&\multicolumn{1}{c}{\shortstack{Birth weight\\(in g)}}&\multicolumn{1}{c}{BMI}&\multicolumn{1}{c}{\shortstack{Height\\(in cm)}}&\multicolumn{1}{c}{\shortstack{Cardiovascular\\Disease}}&\multicolumn{1}{c}{\shortstack{Type II\\Diabetes}}&\multicolumn{1}{c}{\shortstack{Sugar\\intake}}&\multicolumn{1}{c}{\shortstack{Carbohydrate\\intake}}&\multicolumn{1}{c}{\shortstack{Fats\\intake}}\\
\midrule
\textbf{Panel A: By trimester}&                     &                     &                     &                     &                     &                     &                     &                     &                     \\
\addlinespace
E: \textit{In utero} during trimester 1&      -0.187         &      47.413         &       0.131         &       0.268         &       0.014         &       0.001         &      -0.003         &      -0.005         &       0.003         \\
            &     (0.224)         &    (35.995)         &     (0.175)         &     (0.275)         &     (0.022)         &     (0.011)         &     (0.005)         &     (0.006)         &     (0.005)         \\
\addlinespace
E: \textit{In utero} during trimester 2&       0.054         &      62.213\sym{***}&      -0.171         &       0.086         &      -0.005         &      -0.005         &      -0.003         &      -0.005         &       0.002         \\
            &     (0.118)         &    (21.255)         &     (0.123)         &     (0.164)         &     (0.011)         &     (0.005)         &     (0.003)         &     (0.004)         &     (0.003)         \\
\addlinespace
E: \textit{In utero} during trimester 3&       0.207\sym{***}&      -2.570         &      -0.080         &      -0.049         &      -0.005         &      -0.002         &      -0.002         &      -0.000         &      -0.002\sym{*}  \\
            &     (0.071)         &    (12.630)         &     (0.061)         &     (0.084)         &     (0.008)         &     (0.003)         &     (0.001)         &     (0.002)         &     (0.001)         \\
\midrule
\textbf{Panel B: De-rationing until May '49}&                     &                     &                     &                     &                     &                     &                     &                     &                     \\
\addlinespace
E: \textit{In utero} during de-rationing&       0.137\sym{***}&      16.512\sym{*}  &      -0.082\sym{**} &       0.042         &      -0.002         &      -0.002         &      -0.002\sym{*}  &      -0.001         &      -0.001         \\
            &     (0.048)         &     (8.952)         &     (0.041)         &     (0.061)         &     (0.005)         &     (0.002)         &     (0.001)         &     (0.001)         &     (0.001)         \\
\midrule
\textbf{Panel C: De-rationing until May '49 \& by trimester}&                     &                     &                     &                     &                     &                     &                     &                     &                     \\
\addlinespace
E: \textit{In utero} during trimester 1&       0.144         &      19.664         &      -0.116         &       0.254\sym{**} &      -0.000         &      -0.001         &      -0.001         &       0.003         &      -0.001         \\
            &     (0.102)         &    (16.360)         &     (0.083)         &     (0.103)         &     (0.009)         &     (0.004)         &     (0.003)         &     (0.003)         &     (0.002)         \\
\addlinespace
E: \textit{In utero} during trimester 2&       0.168\sym{*}  &      22.929\sym{*}  &      -0.024         &      -0.160         &       0.004         &      -0.002         &      -0.004\sym{**} &      -0.003         &       0.002         \\
            &     (0.094)         &    (13.635)         &     (0.081)         &     (0.108)         &     (0.009)         &     (0.004)         &     (0.002)         &     (0.002)         &     (0.002)         \\
\addlinespace
E: \textit{In utero} during trimester 3&       0.114         &      10.929         &      -0.097\sym{*}  &       0.040         &      -0.008         &      -0.003         &      -0.001         &      -0.002         &      -0.003\sym{*}  \\
            &     (0.077)         &    (17.254)         &     (0.056)         &     (0.090)         &     (0.008)         &     (0.003)         &     (0.001)         &     (0.002)         &     (0.001)         \\
\midrule
No. of observations&       84184         &       47476         &       84554         &       84641         &       84672         &       84637         &       36131         &       36131         &       36131         \\
 \\
\bottomrule
\addlinespace[.75ex]
\end{tabular}
}
\label{tab:reg3}
\caption*{\noindent\scriptsize Estimates from Equation (\ref{eq:E}). Panel A reports the estimates of exposure to derationing on the different outcomes of interest, distinguishing by the trimester of exposure. Exposure in the first trimester is defined as those born between December 24, 1949 and May 12, 1950. Those born between September 24, 1949 and December 23, 1949 are exposed in the second trimester, and those born between April 24, 1949 and September 23, 1949 are exposed in the third trimester.  Panel B defines the derationing period to be between 24 April 1949 and 31 May 1949, following the literature that suggests that many shops had run out of confectionery by then. Panel C distinguishes between trimesters of exposure within this shorter period of derationing. First trimester exposure is here defined as those born between November 13, 1949 and February 28, 1950. Those born between August 13, 1949 and November 12, 1949 are exposed in the second trimester, and those born between April 24, 1949 and August 12, 1949 are exposed in the third trimester. All regressions include administrative county of birth fixed effects. Robust standard errors, clustered by administrative county of birth, in parentheses. * $p < 0.10$, ** $p < 0.05$, *** $p < 0.01$.}
\end{table}

\begin{table}[H]
\caption{Only including those already in utero at the time of the derationing announcement}
\centering
{\tiny
\begin{tabular}{lcccccccccccccccccccc}
\toprule
            &\multicolumn{1}{c}{(1)}&\multicolumn{1}{c}{(2)}&\multicolumn{1}{c}{(3)}&\multicolumn{1}{c}{(4)}&\multicolumn{1}{c}{(5)}&\multicolumn{1}{c}{(6)}&\multicolumn{1}{c}{(7)}&\multicolumn{1}{c}{(8)}&\multicolumn{1}{c}{(9)}\\
            &\multicolumn{1}{c}{\shortstack{Years of\\Education}}&\multicolumn{1}{c}{\shortstack{Birth weight\\(in g)}}&\multicolumn{1}{c}{BMI}&\multicolumn{1}{c}{\shortstack{Height\\(in cm)}}&\multicolumn{1}{c}{\shortstack{Cardiovascular\\Disease}}&\multicolumn{1}{c}{\shortstack{Type II\\Diabetes}}&\multicolumn{1}{c}{\shortstack{Sugar\\intake}}&\multicolumn{1}{c}{\shortstack{Carbohydrate\\intake}}&\multicolumn{1}{c}{\shortstack{Fats\\intake}}\\
\midrule
\textbf{Panel A: No area fixed effects}&                     &                     &                     &                     &                     &                     &                     &                     &                     \\
\addlinespace
E: \textit{In utero} during derationing&       0.169\sym{***}&      12.767         &      -0.098\sym{*}  &      -0.040         &      -0.004         &      -0.003         &      -0.002\sym{*}  &      -0.002         &      -0.001         \\
            &     (0.061)         &    (10.717)         &     (0.056)         &     (0.073)         &     (0.005)         &     (0.002)         &     (0.001)         &     (0.001)         &     (0.001)         \\
\midrule
\textbf{Panel B: Administrative county FE}&                     &                     &                     &                     &                     &                     &                     &                     &                     \\
\addlinespace
E: \textit{In utero} during derationing&       0.143\sym{**} &      13.583         &      -0.085\sym{*}  &      -0.055         &      -0.004         &      -0.003         &      -0.002\sym{*}  &      -0.002         &      -0.001         \\
            &     (0.064)         &    (12.151)         &     (0.047)         &     (0.074)         &     (0.007)         &     (0.002)         &     (0.001)         &     (0.001)         &     (0.001)         \\
\midrule
No. of observations&       76772         &       43232         &       77110         &       77191         &       77228         &       77196         &       32933         &       32933         &       32933         \\
 \\
\bottomrule
\addlinespace[.75ex]
\end{tabular}
}
\label{tab:timing}
\caption*{\noindent\scriptsize Sample includes those born between April 1947 and 24 November 1949 (i.e., 9 months post the announcement of derationing), and between 13 May 1950 (i.e., 9 months after re-rationing) and 12 May 1952, with those born between 24 April 1949 and 24 November 1949 defined as treated. Panel A does not control for region of birth fixed effects; Panel B controls for administrative county fixed effects. Robust standard errors, clustered by year-month of birth in Panel A and by administrative county of birth in Panel B, in parentheses. * $p < 0.10$, ** $p < 0.05$, *** $p < 0.01$.}
\end{table}


\begin{table}[H]
\caption{Robustness analysis - allowing for different specifications of the calendar time trend effect}
\centering
{\tiny
\begin{tabular}{lcccccccccccccccccccc}
\toprule
            &\multicolumn{1}{c}{(1)}&\multicolumn{1}{c}{(2)}&\multicolumn{1}{c}{(3)}&\multicolumn{1}{c}{(4)}&\multicolumn{1}{c}{(5)}&\multicolumn{1}{c}{(6)}&\multicolumn{1}{c}{(7)}&\multicolumn{1}{c}{(8)}&\multicolumn{1}{c}{(9)}\\
            &\multicolumn{1}{c}{\shortstack{Years of\\Education}}&\multicolumn{1}{c}{\shortstack{Birth weight\\(in g)}}&\multicolumn{1}{c}{BMI}&\multicolumn{1}{c}{\shortstack{Height\\(in cm)}}&\multicolumn{1}{c}{\shortstack{Cardiovascular\\Disease}}&\multicolumn{1}{c}{\shortstack{Type II\\Diabetes}}&\multicolumn{1}{c}{\shortstack{Sugar\\intake}}&\multicolumn{1}{c}{\shortstack{Carbohydrate\\intake}}&\multicolumn{1}{c}{\shortstack{Fats\\intake}}\\
\midrule
\textbf{Panel A: One linear trend in year-month of birth}&                     &                     &                     &                     &                     &                     &                     &                     &                     \\
\addlinespace
E: \textit{In utero} during derationing&       0.123\sym{***}&      -1.920         &      -0.062         &       0.097\sym{*}  &      -0.001         &      -0.002         &      -0.001         &      -0.001         &      -0.001         \\
            &     (0.039)         &     (8.622)         &     (0.040)         &     (0.059)         &     (0.004)         &     (0.002)         &     (0.001)         &     (0.001)         &     (0.001)         \\
\midrule
\textbf{Panel B: One quadratic trend in year-month of birth}&                     &                     &                     &                     &                     &                     &                     &                     &                     \\
\addlinespace
E: \textit{In utero} during derationing&       0.082         &      -7.521         &      -0.069         &       0.084         &      -0.001         &      -0.001         &      -0.002         &      -0.000         &      -0.001         \\
            &     (0.053)         &    (10.280)         &     (0.052)         &     (0.070)         &     (0.006)         &     (0.002)         &     (0.001)         &     (0.001)         &     (0.001)         \\
\midrule
\textbf{Panel C: Differential quadratic trends for treated and control}&                     &                     &                     &                     &                     &                     &                     &                     &                     \\
\addlinespace
E: \textit{In utero} during derationing&       0.128\sym{**} &      13.561         &      -0.093         &       0.010         &      -0.004         &      -0.001         &      -0.003\sym{***}&      -0.002         &       0.000         \\
            &     (0.063)         &    (11.508)         &     (0.058)         &     (0.073)         &     (0.007)         &     (0.003)         &     (0.001)         &     (0.001)         &     (0.001)         \\
\midrule
No. of observations&       84184         &       47476         &       84554         &       84641         &       84672         &       84637         &       36131         &       36131         &       36131         \\
 \\
\bottomrule
\addlinespace[.75ex]
\end{tabular}
}
\label{tab:difftrend}
\caption*{\noindent\scriptsize Estimates from Equation (\ref{eq:E}), with the treatment dummy ($E$) specified as those born between 24 April 1949 and 12 May 1950. Panel A accounts for a linear function in the year-month of birth; Panel B allows for a quadratic function in the year-month of birth; Panel C allows for different quadratic functions for exposed and non-exposed. All regressions control for administrative-county-of-birth fixed effects. Robust standard errors, clustered by administrative county, in parentheses. * $p < 0.10$, ** $p < 0.05$, *** $p < 0.01$.}
\end{table}


\begin{table}[H]
\caption{Gender differences}
\centering
{\tiny
\begin{tabular}{lcccccccccccccccccccc}
\toprule
            &\multicolumn{1}{c}{(1)}&\multicolumn{1}{c}{(2)}&\multicolumn{1}{c}{(3)}&\multicolumn{1}{c}{(4)}&\multicolumn{1}{c}{(5)}&\multicolumn{1}{c}{(6)}&\multicolumn{1}{c}{(7)}&\multicolumn{1}{c}{(8)}&\multicolumn{1}{c}{(9)}\\
            &\multicolumn{1}{c}{\shortstack{Years of\\Education}}&\multicolumn{1}{c}{\shortstack{Birth weight\\(in g)}}&\multicolumn{1}{c}{BMI}&\multicolumn{1}{c}{\shortstack{Height\\(in cm)}}&\multicolumn{1}{c}{\shortstack{Cardiovascular\\Disease}}&\multicolumn{1}{c}{\shortstack{Type II\\Diabetes}}&\multicolumn{1}{c}{\shortstack{Sugar\\intake}}&\multicolumn{1}{c}{\shortstack{Carbohydrate\\intake}}&\multicolumn{1}{c}{\shortstack{Fats\\intake}}\\
\midrule
\textbf{Panel A: Women}&                     &                     &                     &                     &                     &                     &                     &                     &                     \\
\addlinespace
D: \textit{In utero} during de-rationing&       0.091         &      22.253\sym{*}  &      -0.106         &      -0.002         &       0.004         &      -0.002         &      -0.001         &      -0.001         &      -0.001         \\
            &     (0.068)         &    (12.486)         &     (0.064)         &     (0.087)         &     (0.006)         &     (0.002)         &     (0.001)         &     (0.002)         &     (0.002)         \\
\midrule
Mean of outcome&       13.93         &     3331.85         &       27.58         &      168.51         &        0.33         &        0.05         &        0.23         &        0.48         &        0.33         \\
No. of observations&       45992         &       29024         &       46209         &       46251         &       46254         &       46260         &       19016         &       19016         &       19016         \\
\midrule
\textbf{Panel B: Men}&                     &                     &                     &                     &                     &                     &                     &                     &                     \\
\addlinespace
D: \textit{In utero} during de-rationing&       0.228\sym{**} &       4.254         &      -0.049         &       0.004         &      -0.012         &      -0.003         &      -0.003\sym{*}  &      -0.002         &      -0.001         \\
            &     (0.090)         &    (15.570)         &     (0.062)         &     (0.092)         &     (0.008)         &     (0.004)         &     (0.001)         &     (0.002)         &     (0.002)         \\
\midrule
Mean of outcome&       13.93         &     3331.85         &       27.58         &      168.51         &        0.33         &        0.05         &        0.23         &        0.48         &        0.33         \\
No. of observations&       38192         &       18452         &       38345         &       38390         &       38418         &       38377         &       17115         &       17115         &       17115         \\
 \\
\bottomrule
\addlinespace[.75ex]
\end{tabular}
}
\label{tab:reg4}
\caption*{\noindent\scriptsize Estimates from Equation (\ref{eq:E}), with the treatment dummy ($E$) specified as those born between 24 April 1949 and 12 May 1950. Panel A shows the results for women; Panel B shows the results for men. All regressions include administrative county of birth fixed effects. Robust standard errors, clustered by adminsitrative county of birth, in parentheses. * $p < 0.10$, ** $p < 0.05$, *** $p < 0.01$.}
\end{table}
\end{landscape}


\renewcommand\thepage{A\arabic{page}}
\setcounter{page}{1}

\newpage
\appendix 


\setcounter{table}{0} 
\setcounter{figure}{0} 
\setcounter{equation}{0} 
\renewcommand{\thetable}{\Alph{section}A.\arabic{table}}
\renewcommand{\thefigure}{\Alph{section}A.\arabic{figure}}
\renewcommand{\theequation}{\Alph{section}A.\arabic{equation}}
\renewcommand{\thesubsection}{Appendix A}
\renewcommand{\thesubsubsection}{A.\arabic{subsubsection}}

\Large{\noindent \textbf{For Online Publication}}

\normalsize{
\subsection{Additional detail on the UK rationing system} \label{sec:AppendixA}
}

Before the Second World War, Britain imported 55m tons of food per year, including 50\% of meats, 70-80\% of its cheese and sugar, nearly 80\% of fruits, about 70\% of cereals and fats, and 90\% of butter, cereal and flour (Ministry of Food, 1951). A month after the war had started, this had dropped to 12m on an annual base. 

Five days after the war was declared, in September 1939, individuals received their ration books. Households had to register with a local supplier; the details of which were stamped in the book, as one could only purchase the ration from the registered supplier. The ration books were not used until four months later (on January 8, 1940). The ration coupon was necessary (and a government guarantee) for the individual to get their share of the item. The coupons could not be used on other produce, nor could they be carried forward. As such, rationing was not applied to e.g. seasonal items, as the government was unable to guarantee their year-round supply. Other items that the government did not want to ration include fresh fruit and vegetables, bread, potatoes, cigarettes and beer. One reason was that supplies were limited anyway (e.g., fresh fruit) so the government could not guarantee its availability, or the government wanted to provide sufficient quantities of cheap energy (i.e., bread, potatoes). Another was that rationing these foods would impact morale, which the government wanted to avoid, and -- for tobacco -- it would have conflicted with its revenue-raising function (Zweiniger-Bargielowska, 2000).\footnote{The ``Dig for Victory'' campaign meant that the consumption of homegrown fruit and vegetables from individuals' allotments also increased during the war. Furthermore, daylight savings were put ahead by two hours every March, ensuring more daylight hours available for farming and gardening.} In summary, rationing was based on the principle that all essential nutrients, even those in short supply, should be equally available to everyone, to the extent necessary to maintain health, and at controlled prices (Ministry of Health, 1946).

There were different ration books for different population groups. For example, most people received the standard blue ration book, but pregnant women, nursing mothers and children under five received green ones, which entitled them to additional milk, eggs and meat, orange juice, cod liver oil, and vitamin A and D (Davies, 2014). ``Points rationing'' was introduced in December 1941 via a separate pink rationing book. As explained in Section 2, it ran in conjunction with regular coupon rationing, but was more flexible, giving shoppers greater choice. It could also be used for purchases at grocers other than the one the household had registered with. However, points rationing was mainly used for tinned goods. The Ministry of Food could adjust the price (i.e. the number of points that an item would cost) in order to try to achieve a balance between supply and demand. In general, prices of unrationed goods were strictly controlled as well by the government. As a result, unrationed goods were not always available.

\autoref{tab:Timeline}  shows a rough timeline of when some of the main foods came on and off rationing. The amount of food that one could get for one coupon tended to vary somewhat over time in line with the quantity of food imports and home-grown produce; see Zweiniger-Bargielowska (2000) for detailed information.

There were other food-related schemes in operation before, during and after the War. For example, the National Milk Scheme provided milk to vulnerable individuals (e.g., children, expectant and nursing mothers). The Milk in Schools Scheme had already been in place since October 1934, providing milk to many (state) school children. The milk provided through these schemes increased somewhat during the war years but there were no large changes to the eligibility requirements.

\begin{table}[H]
\caption{Start and end dates of rationing in the UK for selected main foods}
\centering
{\scriptsize
\begin{tabular}{lrrlccc}
\toprule
			                &    Start 		 & 	 End 	&   Notes 									 \\
\midrule
National Registration Day 	&    			 &  		   &  Sep 29 1939: Provide details of all household members \\
Gammon, ham 				&     Jan 8, 1940 &  Oct, 3 1952 &   \\
Bacon						&     Jan 8, 1940 &  Jul, 3 1954 &   \\
Butter 						&     Jan 8, 1940 &  May, 8 1954 &   \\
Sugar 						&     Jan 8, 1940 &  Sep, 27 1953 &   \\
Meat (incl beef, veal, mutton, pork) &    Mar, 11 1940 &  Jul, 4 1954 &   \\
Margarine 					&     Jul 1940 &  May 8, 1954 &   \\
Tea 						&     Jul 1940 &  Oct 5, 1952 &   \\
Cooking fat 				&     Jul 1940 &  May 8, 1954 &   \\
Cheese 						& 	  May 5, 1941 &    May 1954 &  Vegetarians were allowed an extra 85g on top \\
							& 				 & 			   &  of the standard ration \\
Eggs 						& 	    Jun 1941 & Mar 26, 1953 &   \\
Milk 						&       Nov 1941 &        1950 &   \\
Tinned veg, breakfast cereals 				&     Feb 9, 1942 & May 19, 1950 &  Points rationing \\
Dried and canned fruit, rice& 	    Jan 1942 & May 19, 1950 &  Points rationing \\
Biscuits 					& 	    Aug 1942 & May 19, 1950 &  Points rationing  \\
Oats (flaked and rolled) 	&       Dec 1942 & May 19, 1950 &  Points rationing  \\
Syrup, treacle 				&       Jun 1942 & May 19, 1950 &  Transferred from standard rationing to points scheme \\
Confectionery 				& 	 Jul 26, 1942 &  Feb 5, 1953 &  Temporary derationing between April 24 - August 13 1949 \\
\bottomrule
\addlinespace[.75ex]
\end{tabular}
}
\label{tab:Timeline}
\caption*{\noindent\scriptsize Notes: The precise date of introduction and removal of rationing is not always known. }
\end{table}


\section*{References}

\begin{description}

\item Davies, N. (2014), Europe: A history, Random House.

\item Ministry of Food (1951), WWII National Food Survey, 1951. Rationing statistics: Straight rationing (Part II), 1948-1951. MAF 156/263. The National Archives.

\item Ministry of Health (1946), On the state of the public health during six years of war. London, His Majesty's Stationery Office.

\item Zweiniger-Bargielowska, I. (2000), Austerity in Britain: Rationing, Controls, and Consumption, 1939-1955, Oxford University Press.

\end{description}


\newpage
\setcounter{table}{0} 
\setcounter{figure}{0} 
\setcounter{equation}{0} 
\renewcommand{\thetable}{\Alph{section}B.\arabic{table}}
\renewcommand{\thefigure}{\Alph{section}B.\arabic{figure}}
\renewcommand{\theequation}{\Alph{section}B.\arabic{equation}}
\renewcommand{\thesubsection}{Appendix B}
\renewcommand{\thesubsubsection}{B.\arabic{subsubsection}}

\subsection{Genetic Analysis} \label{sec:AppendixB}
\subsubsection{An introduction to genetics}
The human genome consists of over 3 billion base pairs (6 billion bases) in each cell nucleus, with four possible bases: adenine (A), thymine (T), guanine (G) and cytosine (C).\footnote{A base pair is set of two bases, with A always pairing with T, and C always pairing with G.} Over 99\% of the human genome is identical between any two unrelated human beings. The remaining $<$1\% differs between individuals, with a Single Nucleotide Polymorphism (or SNP, pronounced ‘snip’) being the most common form of genetic variation. A SNP is a one base-pair substitution at a particular location (locus) on the human genome.

To identify genetic variants that are associated with a particular trait of interest, such as years of education or BMI, so-called Genome-Wide Association Studies (GWAS) relate each SNP to the trait in a hypothesis free-approach. This is done by a series of OLS regressions of the trait on the genotype of a single SNP and a set of covariates (normally sex, age, genotype array, and the first set of principal components of the genetic data to account for ancestry). GWAS have shown that most human traits are highly polygenic, meaning that they are affected by many SNPs, each with very small effect sizes. To increase their predictive power, they can then be aggregated into so-called polygenic scores (also known as polygenic indices), as shown in \autoref{eq:PGS}. Here, it is important that the $\beta_j$'s, the effect sizes of each SNP $j$, are obtained from an independent GWAS, as not doing so leads to overfitting. 

GWAS sample sizes have grown substantially in recent years, meaning that (a) SNPs with very small effect sizes are more likely to be identified, and (b) that the effect sizes are estimated with increased precision. Indeed, we have seen large improvements in genetic prediction, with initial GWASs being able to explain less than 1\% of the variation in the trait of interest (e.g., years of education), to more recent ones explaining over 10\% of its variation (see e.g., Okbay et al., 2022).

\subsubsection{Our tailor-made GWAS}
The full UK Biobank data release contains all successfully genotyped samples (n=488,377). 
Pre-imputation QC, phasing and imputation are described in detail in Bycroft et al. (2018). In brief, prior to phasing, we removed multiallelic SNPs or those with a minor allele frequency (MAF) $\leq$1\%. Phasing of genotype data was performed using a modified version of the SHAPEIT2 algorithm (O'Connell et al., 2016). Genotype imputation to a reference set combining the UK10K haplotype and HRC reference panels (Huang et al., 2015) was performed using IMPUTE2 algorithms (Howie et al., 2011). Our analysis is restricted to autosomal variants within the HRC site list using a graded filtering with varying imputation quality for different allele frequency ranges. 

We follow the literature and exclude individuals with sex-mismatch (derived by comparing genetic sex and reported sex) or individuals with sex-chromosome aneuploidy from the analysis (n=814). Furthermore, we restrict the sample to individuals of white British ancestry who self-report as `White British' and who have very similar ancestral backgrounds according to the PCA (n=409,703), as described by Bycroft et al. (2018). Finally, our GWAS discovery sample only includes unrelated individuals; we exclude genetically related individuals as well as our analysis sample (i.e., those born between 24 April 1947 and 13 May 1952) before running our GWAS.

Depending on the distribution of the outcome, we use a linear or logistic regression in the genome-wide association study, using the software PLINKv2.00. We specify the genotype array, gender and the first 10 principal components as covariates.

\subsubsection{Calculating Polygenic Scores}
We use LDpred (Vilhj{\'a}lmsson et al., 2015) to construct all polygenic scores. This Python-based software package adjusts GWAS summary statistics for the effects of linkage disequilibrium (LD). It uses a Bayesian method, assuming a prior for the genetic architecture of the trait of interest in combination with information on the linkage disequilibrium from an external reference panel (30,000 unrelated individuals in UK Biobank) to estimate posterior mean effect sizes from summary statistics. We construct the polygenic scores using an LD window of M/3000 where M is the total number of SNPs. We assume the fraction of SNPs with non-zero effects to be 0.3. Our results are robust to taking a prior of 1 instead of 0.3. We restrict the analysis to 1.2M HapMap 3 SNPs (International HapMap3 Consortium, 2010) to ensure a set of common, well-imputed variants.


\section*{References}

\begin{description}

\item Bycroft, C., Freeman, C., Petkova, D., ..., Marchini, J. (2018), The UK Biobank resource with deep phenotyping and genomic data, {\em Nature}  562, 203--209.

\item Howie, B., Marchini, J. and Stephens, M. (2011), Genotype Imputation with Thousands of Genomes, {\em Genes|Genomes|Genetics} 1, 457--470.

\item Huang, J., Howie, B., McCarthy, S., ..., Soranzo, N. (2015), Improved imputation of low-frequency and rare variants using the UK10K haplotype reference panel, {\em Nature Communications} 6, 8111.

\item International HapMap 3 Consortium (2010), Integrating common and rare genetic variation in diverse human populations, {\em Nature} 467, 52--58.

\item O'Connell, J., Sharp, K., Shrine, N., ..., Marchini, J. (2016), Haplotype estimation for biobank-scale data sets, {\em Nature Genetics} 48, 817--820. 

\item Okbay, A., Wu, Y., ..., Young, A. (2022), Polygenic prediction of educational attainment within and between families from genome-wide association analyses in 3 million individuals. {\em Nature Genetics} 54, 437--449

\item Vilhj{\'a}lmsson, B. J., Yang, J., Finucane, H. K., ..., Hayeck, T. (2015), Modeling linkage disequilibrium increases accuracy of polygenic risk scores, {\em The American Journal of Human Genetics} 97.4, 576--592.

\end{description}


\newpage
\setcounter{table}{0} 
\setcounter{figure}{0} 
\setcounter{equation}{0} 
\renewcommand{\thetable}{\Alph{section}C.\arabic{table}}
\renewcommand{\thefigure}{\Alph{section}C.\arabic{figure}}
\renewcommand{\theequation}{\Alph{section}C.\arabic{equation}}
\renewcommand{\thesubsection}{Appendix C}
\renewcommand{\thesubsubsection}{C.\arabic{subsubsection}}

\subsection{Additional Figures}\label{sec:AppendixC}

\begin{figure}[H]
\caption{Variation in the sweets ration; 1942 -- 1948}
\label{fig:rations}
  \centering
  \includegraphics[width=1\linewidth]{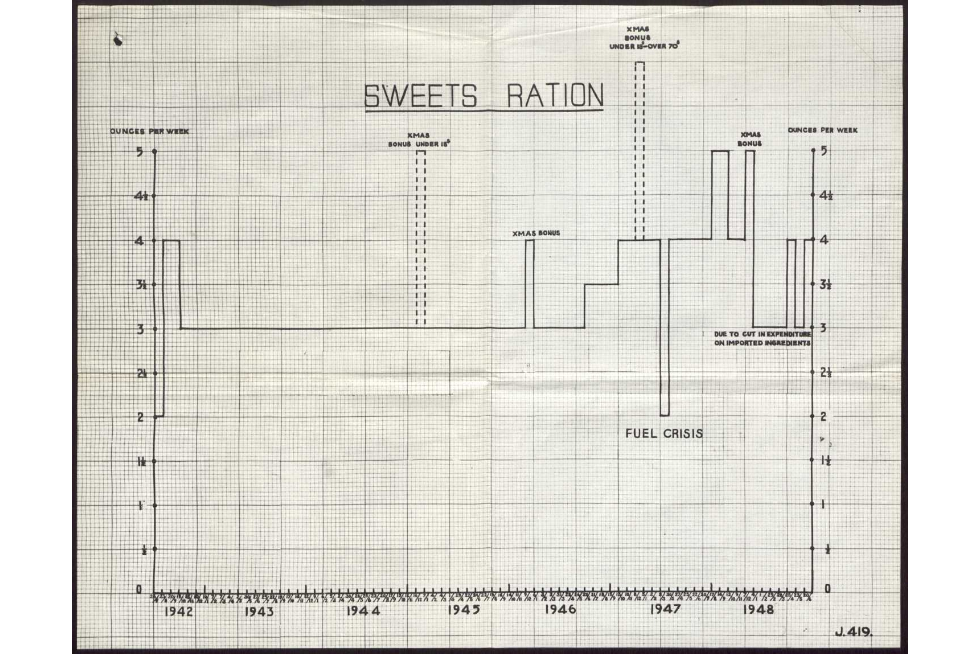}
\caption*{\scriptsize{The figure shows the variation in the sweets ration between its introduction on July 26, 1942 and the end of 1948. Source: MAF 156/263, 1951.}}
\end{figure}

\begin{figure}[H]
\caption{Gene-environment correlations}
\centering
\label{fig:rGE}
\begin{subfigure}{.4\linewidth}
  \centering
  \includegraphics[width=1\linewidth]{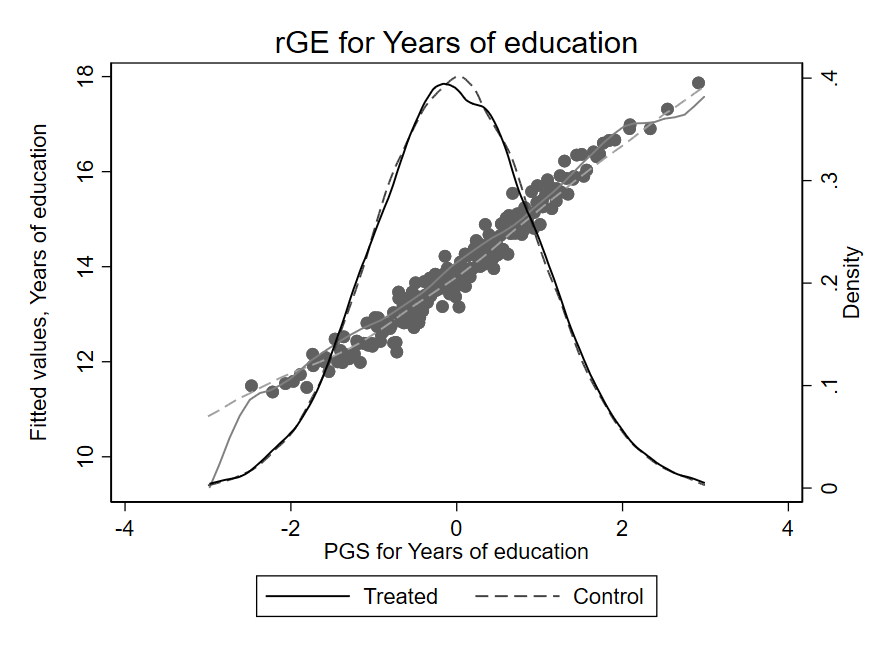}
  \caption*{}
\end{subfigure}
\begin{subfigure}{.4\linewidth}
  \centering
  \includegraphics[width=1\linewidth]{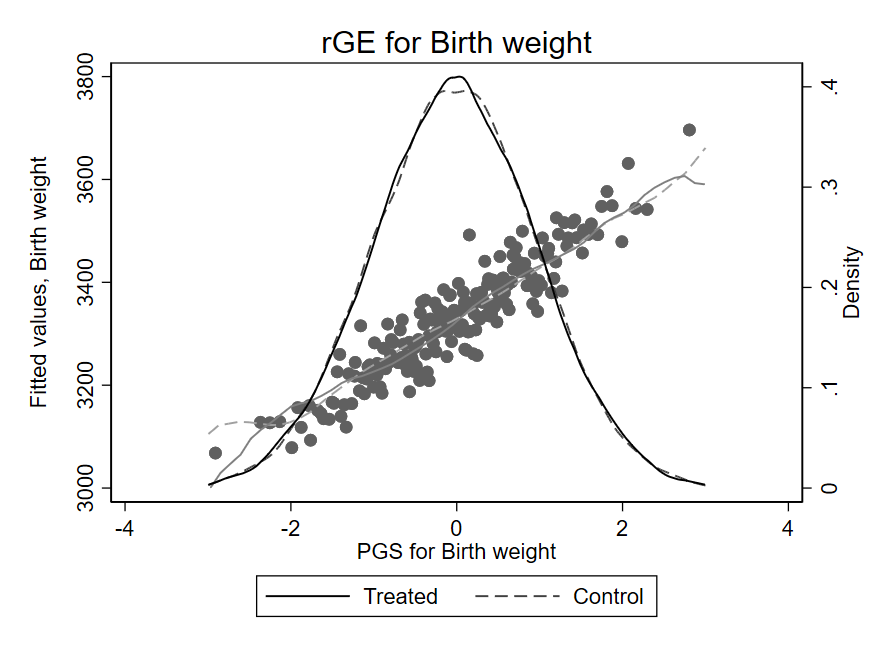}
  \caption*{}
\end{subfigure}\\[1ex]
\begin{subfigure}{.4\linewidth}
  \centering
  \includegraphics[width=1\linewidth]{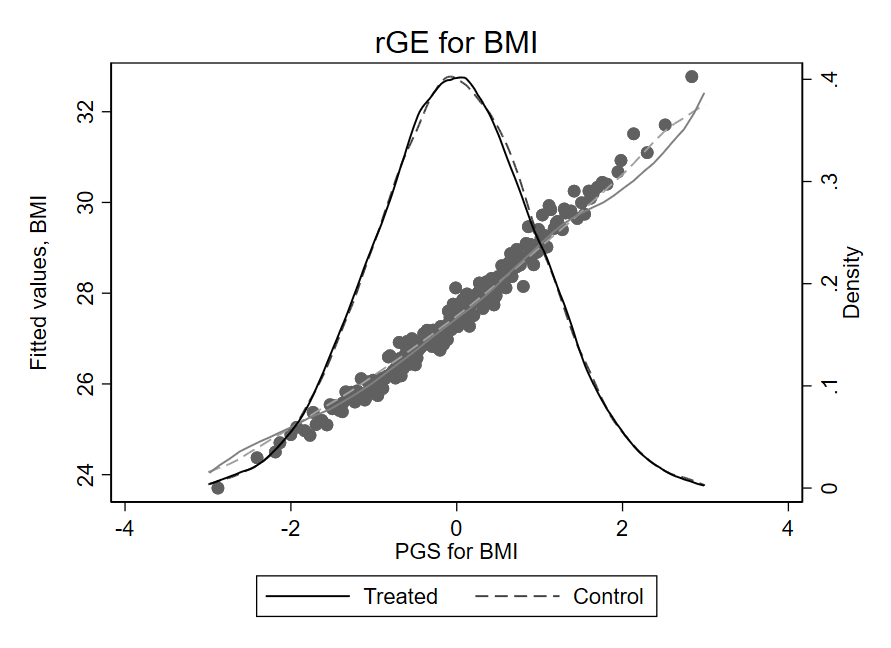}
  \caption*{}
\end{subfigure}
\begin{subfigure}{.4\linewidth}
  \centering
  \includegraphics[width=1\linewidth]{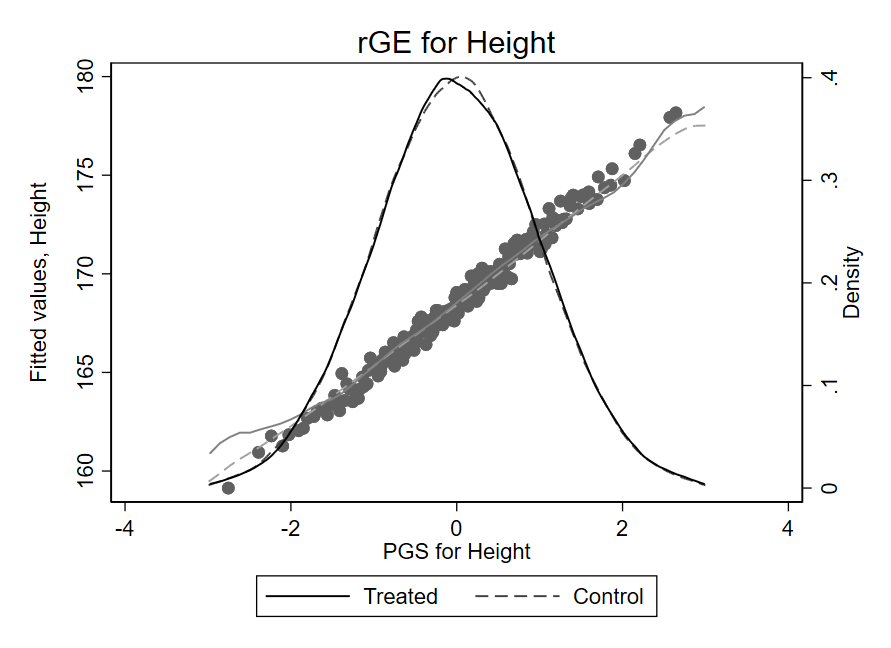}
  \caption*{}
\end{subfigure}\\[1ex]
\begin{subfigure}{.4\linewidth}
  \centering
  \includegraphics[width=1\linewidth]{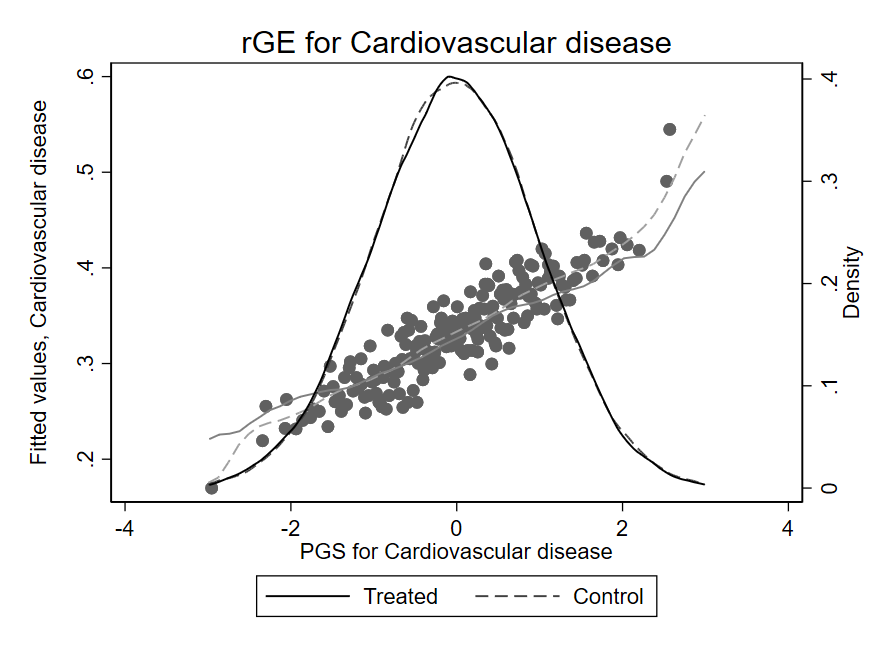}
  \caption*{}
\end{subfigure}
\begin{subfigure}{.4\linewidth}
  \centering
  \includegraphics[width=1\linewidth]{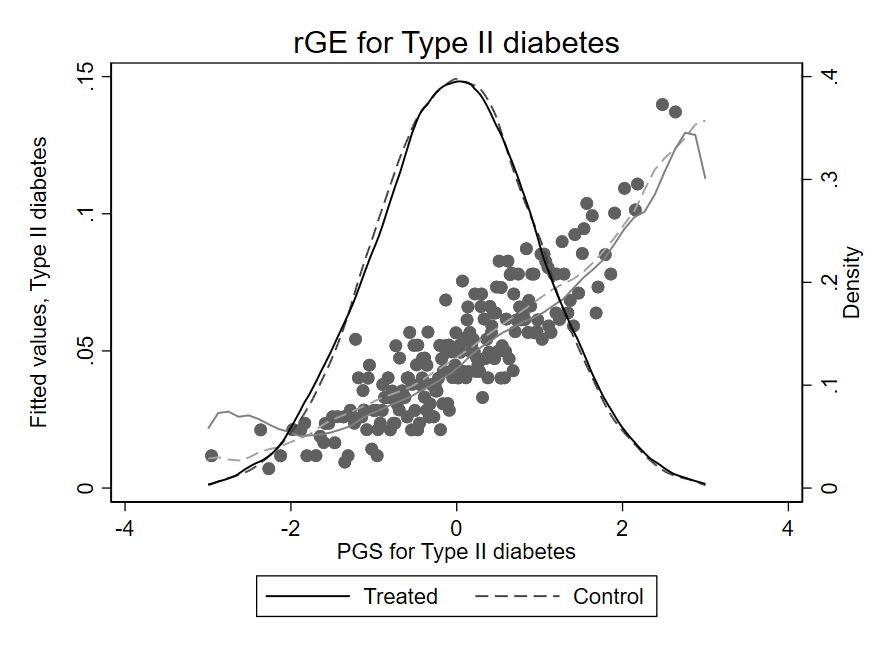}
  \caption*{}
\end{subfigure}\\[1ex]
\begin{subfigure}{.4\linewidth}
  \centering
  \includegraphics[width=1\linewidth]{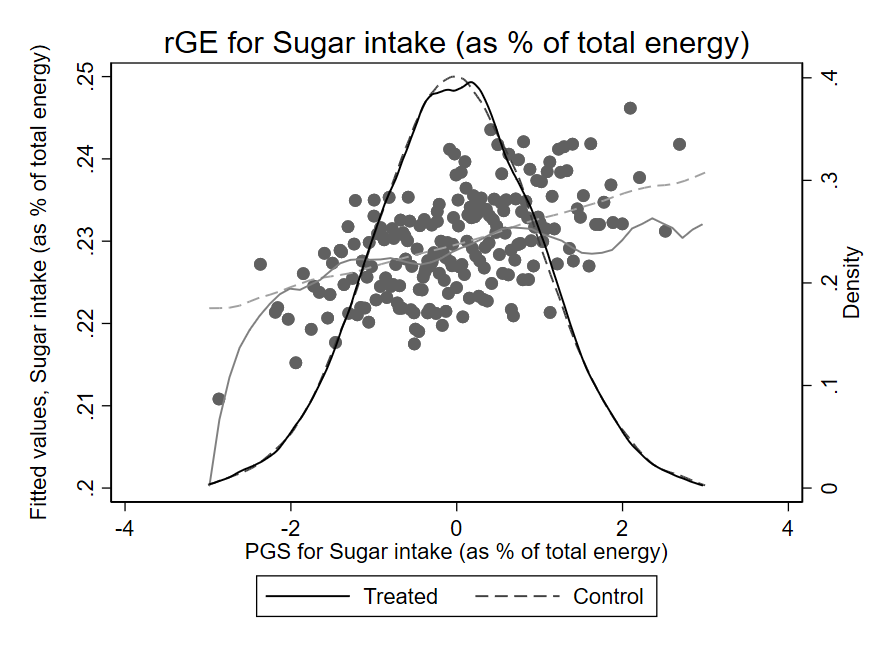}
  \caption*{}
\end{subfigure}
\begin{subfigure}{.4\linewidth}
  \centering
  \includegraphics[width=1\linewidth]{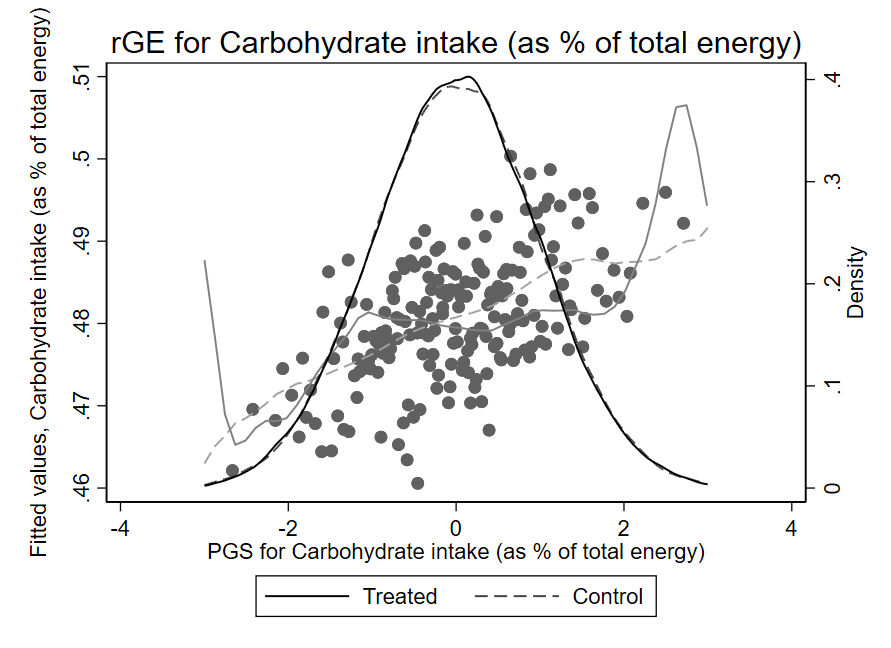}
  \caption*{}
\end{subfigure}\\[1ex]
\caption*{\scriptsize{Notes: The figures plot the distribution of the different polygenic scores by treatment status and show the predictive power of the polygenic score (created using our tailor-made GWAS on the non-sibling sample of the UK Biobank, excluding the estimation sample) for their respective outcomes.}}
\end{figure}


\end{document}